\documentclass[10pt,twocolumn,letterpaper]{article}

\usepackage[T1]{fontenc}
\usepackage{newtxtext}

\usepackage{amsmath}
\usepackage{amsthm}
\usepackage{newtxmath}

\usepackage{helvet}
\usepackage{courier}

\usepackage[
    letterpaper,
    left=0.75in,
    right=0.75in,
    top=0.75in,
    bottom=1.00in,
    columnsep=0.375in
]{geometry}

\usepackage[hyphens]{url}
\usepackage{graphicx}
\usepackage{caption}
\usepackage{natbib}
\usepackage{threeparttable}
\usepackage{makecell}
\usepackage{booktabs}
\usepackage{algorithm}
\usepackage{algorithmic}
\usepackage{multirow}
\usepackage{placeins}
\usepackage{microtype}
\usepackage{titlesec}
\usepackage[hidelinks]{hyperref}

\newtheorem{Definition}{Definition}

\titleformat{\section}
    {\large\bfseries\centering}
    {}
    {0pt}
    {}

\titleformat{\subsection}
    {\normalsize\bfseries}
    {}
    {0pt}
    {}

\titleformat{\subsubsection}
    {\normalsize\bfseries}
    {}
    {0pt}
    {}

\titleformat{\paragraph}[runin]
    {\normalsize\bfseries}
    {}
    {0pt}
    {}

\titlespacing*{\section}
    {0pt}
    {2.0ex plus 0.5ex minus 0.2ex}
    {0.8ex}

\titlespacing*{\subsection}
    {0pt}
    {1.6ex plus 0.4ex minus 0.2ex}
    {0.5ex}

\titlespacing*{\subsubsection}
    {0pt}
    {1.2ex plus 0.3ex minus 0.2ex}
    {0.5ex}

\titlespacing*{\paragraph}
    {0pt}
    {1.0ex plus 0.2ex minus 0.1ex}
    {0.5em}

\title{
    \vspace{-1.2em}
    \bfseries
    CHILL-Harness: Counterfactual Harness Learning for
    Efficient Reasoning in Long-Horizon Agents
}

\author{
    Jiarun Fu$^{1}$
    \quad
    Lizhong Ding$^{1,*}$
    \quad
    Sida Chen$^{1}$
    \quad
    Honglei Xin$^{1}$
    \quad
    Chunhui Zhang$^{1}$
    \\[0.35em]
    Pengqi Li$^{1}$
    \quad
    Qiuning Wei$^{1}$
    \quad
    Ye Yuan$^{1}$
    \quad
    Guoren Wang$^{1}$
    \\[0.55em]
    \small
    $^{1}$School of Computer Science and Technology,
    Beijing Institute of Technology, Beijing, China
    \\[0.35em]
    \small
    Contact:
    \href{mailto:jrf@bit.edu.cn}{\texttt{jrfu@bit.edu.cn}}
}

\date{}

\begin{document}

\maketitle

\begin{abstract}
Agent harnesses have become the operational infrastructure of modern
large language model agents, coordinating context, tools, verification, and
execution control to translate latent model capability into reliable
long-horizon behavior.
However, reliable long-horizon behavior requires harness control to adapt to
task demands, execution environments, and evolving execution states, whereas current harnesses predominantly rely on hand-crafted or globally fixed
policies; this mismatch manifests as unnecessary computational overhead and,
in adverse cases, reduced task success.
To address this limitation, we formulate the task of enabling adaptive
orchestration in harness systems as a causal learning problem and propose
\textbf{C}ounterfactual \textbf{H}arness \textbf{I}ntervention
\textbf{L}earning for \textbf{L}ong-Horizon Agents
(\textbf{CHILL-Harness}).
CHILL-Harness intervenes at the orchestration layer to enable
advantage-guided workflow adaptation, thereby improving reasoning and
execution efficiency while preserving task performance.
Specifically, we develop causal intervention effect learning as the
effect-estimation component of CHILL-Harness to estimate intervention-relative workflow advantage from confidence-weighted execution evidence and identify advantageous workflow adaptations.
We further introduce advantage-realizing causal orchestration as its
realization component to adaptively allocate counterfactual reasoning and realize only workflow adjustments supported by sufficient expected
advantage. Finally, we incorporate a success-preserving objective and advantage-margin authorization constraints into CHILL-Harness to promote reliable adaptation. Extensive experiments on heterogeneous long-horizon tasks spanning
information seeking, software engineering, and terminal interaction show that
CHILL-Harness consistently preserves or improves task success while
substantially reducing token consumption and execution time. 

Code:  \href{https://github.com/csdstar/chill-dev}{https://github.com/csdstar/chill-dev}
\end{abstract}

\section{Introduction}

Large language model agents are increasingly deployed on long-horizon tasks \cite{zhou2024webarena,xie2024osworld} that require sustained context management \cite{wang2024agent}, external tool interaction
\cite{yao2024tau}, intermediate verification \cite{pan2024training},
failure recovery \cite{wang2025oscar}, and execution control
\cite{guo2026question}. Agent harnesses have therefore become critical runtime infrastructure for
translating latent model capability into reliable behavior \cite{meng2026agentharness,kapoor2025holisticagentleaderboardmissing}.
However, effective harness control should adapt to task demands
\cite{hu2025evaluating}, execution environments \cite{ba2026ciber}, and evolving execution states \cite{zhang2026stop}, whereas existing harnesses remain largely governed by hand-crafted rules, fixed thresholds, or globally configured workflows
\cite{li2026agentharness,marchand2026quantifyingfrontierllmcapabilities}.
Over long execution horizons, this mismatch can compound redundant reasoning, delayed correction, and unnecessary workflow disruption, increasing execution cost and potentially degrading task success
\cite{he2026memoryarena,sun2026sweworldbuildingsoftwareengineering}.

Existing harness-level approaches to adaptive harness orchestration can
be broadly grouped into two categories: (1) task-oriented engineered harnesses such as Terminus-KIRA integrate hand-engineered reasoning and tool-use procedures
across long-horizon tasks \cite{terminuskira2026}, OpenHands and
CodeSweep--SWE-agent provide iterative repository inspection, editing, and
verification workflows for software repair
\cite{wang2024openhands,yang2024sweagent,kimiteam2025kimik2}, and
LemonHarness develops a high-performing execution stack for complex terminal
tasks \cite{ren2026lemonharnesstechnicalreport}; despite their strong
task-level performance, these systems rely primarily on domain-specific or
globally configured control logic and provide limited adaptation to the value
of an intervention at the current execution state; and (2) orchestration and
harness optimization methods improve system-level coordination---AWorld
dynamically organizes multi-agent collaboration
\cite{yu2025aworldorchestratingtrainingrecipe}, OWL Workforce coordinates a
planner, coordinator, and specialized workers through hierarchical
orchestration \cite{hu2025owl}, and Meta-Harness performs outer-loop search
over harness implementations using execution traces and evaluation feedback
\cite{lee2026metaharnessendtoendoptimizationmodel}; however, these methods
primarily optimize collaboration structures or aggregate harness
configurations rather than determine whether a concrete orchestration
intervention causally improves the ongoing workflow.
Consequently, existing methods still lack a unified framework that assesses whether a concrete orchestration adjustment improves the current workflow and translates this assessment into adaptive, success-preserving harness decisions for efficient reasoning.

In this work, we leverage a causal perspective to address this limitation.
Causal reasoning is well suited to decisions whose observed outcomes conflate
the effect of an action with the context in which it is taken
\cite{kuroki2014measurement}, since causal interventions can isolate the
contribution of the action under a fixed context
\cite{rubin1980randomization,pearl2009causality}.
It has therefore been increasingly used to distinguish genuine contributions
from spurious associations in large-language-model reasoning and autonomous
driving
\cite{chi2024unveiling,pourkeshavarz2024cadet,tang2026causalvad}.
Harness orchestration exhibits the same structure: the current workflow
provides a factual reference, while admissible alternatives define
counterfactual interventions under the same execution context.
We therefore formulate the task of enabling adaptive orchestration in harness
systems as a causal learning problem.
Specifically, we introduce the \emph{causal harness orchestration problem}:
\begin{quote}
\emph{How can a harness leverage causal interventions to realize efficient
and admissible workflow orchestration while preserving task success?}
\end{quote}

To address this problem, we propose
\textbf{C}ounterfactual \textbf{H}arness \textbf{I}ntervention
\textbf{L}earning for \textbf{L}ong-Horizon Agents
(\textbf{CHILL-Harness}), a causal framework for adaptive harness
orchestration and efficient long-horizon reasoning.
CHILL-Harness intervenes at the orchestration layer to enable
advantage-guided workflow adaptation.
Specifically, we develop Causal Intervention Effect Learning (CIEL) as its
effect-estimation component, which unifies heterogeneous orchestration
behaviors and estimates their context-conditioned intervention-relative
advantages from confidence-weighted execution evidence.
We further introduce Advantage-Realizing Causal Orchestration (ARCO) as its
realization component, which routes counterfactual deliberation before
candidate generation and authorizes workflow adaptations only when they
exhibit sufficient causal--operational advantage.
Finally, we incorporate a success-preserving objective and advantage-margin authorization constraints to promote reliable orchestration under effect-estimation uncertainty.
Extensive experiments on long-horizon tasks spanning information seeking,
software engineering, and terminal interaction show that CHILL-Harness
preserves or improves task success while substantially reducing token
consumption and execution time.
Our contributions are summarized as follows:
\begin{itemize}
    \item We propose CHILL-Harness, which formulates adaptive orchestration in harness systems as a causal
learning problem over factual and admissible alternative workflows, enabling efficient reasoning through intervention.

    \item We introduce Causal Intervention Effect Learning, which unifies heterogeneous orchestration behaviors and learns their context-conditioned intervention-relative advantages from confidence-weighted execution evidence, enabling selective harness-level intervention. 
    \item We develop Advantage-Realizing Causal Orchestration, which realizes workflow adaptations supported by sufficient estimated advantage, thereby achieving efficient and reliable harness orchestration under success-preserving constraints.
\end{itemize}

\section{Preliminaries and Related Work}
\label{sec:preliminaries_related}
This section introduces the agent-harness setting and causal formulation, followed by a review of adaptive harness orchestration and causal intervention methods.
\subsection{Preliminaries: Causal Adaptation in Harnesses}

An agent harness is the runtime infrastructure that governs how a base
language-model agent maintains context, invokes tools, verifies progress, and
interacts with its task environment
\citep{meng2026agentharness,guo2026question}.
At execution step \(t\), the current context is represented as
\[
\chi_t
=
\left(
s_t,c_t^{\mathrm{env}},g_t
\right),
\]
where \(s_t\), \(c_t^{\mathrm{env}}\), and \(g_t\) denote the execution state,
environment context, and current task objective, respectively.
Given \(\chi_t\), the base agent proposes a factual workflow
\(\omega_t^0\), consisting of the reasoning, tool-use, and execution
operations.

Let \(\mathcal A_t\) denote the admissible workflow set under the current
task, environment, safety, and resource constraints, with
\(\omega_t^0\in\mathcal A_t\), and define
$
\mathcal A_t^{\mathrm{cf}}
=
\mathcal A_t
\setminus
\{\omega_t^0\}.
$
The workflow selected for execution therefore satisfies
$
W_t
\in
\{\omega_t^0\}
\cup
\mathcal A_t^{\mathrm{cf}}.
$
Executing \(W_t=\omega\) produces downstream task performance \(U_t\) under
the common continuation and evaluation protocol.
Adaptive harness orchestration determines whether to retain
\(\omega_t^0\) or execute an admissible alternative under \(\chi_t\).

This decision naturally induces a potential-outcome formulation
\citep{rubin1974estimating,pearl2009causality}.
For each \(\omega\in\mathcal A_t\), let \(U_t(\omega)\) denote the performance
under \(\operatorname{do}(W_t=\omega)\) with \(\chi_t\) fixed.
Under intervention consistency, \(U_t=U_t(W_t)\), and the workflow effect
relative to the factual workflow is
$
\Gamma_t(\omega)
=
\mathbb E
\left[
U_t(\omega)-U_t(\omega_t^0)
\mid
\chi_t
\right],
\Gamma_t(\omega_t^0)=0.
$

\subsection{Related Work: Adaptive Harness Orchestration and Causal
Intervention}

Existing work on adaptive harness orchestration includes engineered execution
systems, automated harness optimization, and component-level adaptation.
OpenHands, SWE-agent, and OWL improve execution through specialized
interfaces or coordination
\citep{wang2024openhands,yang2024sweagent,hu2025owl};
Automated Design of Agentic Systems, Multi-Agent Architecture Search, and
Meta-Harness optimize agent structures, resources, or complete harnesses
\citep{hu2025automated,zhang2025multi,
lee2026metaharnessendtoendoptimizationmodel}, while
Agent Workflow Memory, HiAgent, and OSCAR adapt workflow reuse, context
management, and recovery
\citep{wang2024agent,hu2025hiagent,wang2025oscar}. Causal methods have also been explored for language-model reasoning and agent
decision making.
Causal Sufficiency and Necessity refines reasoning chains through
counterfactual analysis \citep{yu2025causalsufficiency},
Counterfactual Planning revises task-level actions using structural causal
models \citep{fu2026counterfactualplanning}, and
Robust Agents Learn Causal World Models exploits causal environment structure
for robust control \citep{richens2024robust}.

These methods target reasoning steps, agent actions, or environment models, but do not assess heterogeneous harness-level decisions or efficiently realize only workflow-improving adaptations. CHILL-Harness fills this gap by estimating context-conditioned workflow effects and translating them into success-preserving workflow adaptations.

\section{CHILL-Harness: Counterfactual Harness Intervention Learning}
\label{sec:method}

CHILL-Harness decomposes causal harness orchestration into two coupled
problems.

\begin{Definition}[Harness Intervention Effect Problem]
\label{def:harness_intervention_effect}
The Harness Intervention Effect Problem evaluates whether an admissible
workflow alternative improves execution relative to the factual workflow
under the same execution context.
\end{Definition}

\begin{Definition}[Harness Intervention Realization Problem]
\label{def:harness_intervention_realization}
The Harness Intervention Realization Problem determines whether the estimated
benefit of an admissible alternative is sufficient to justify replacing the
factual workflow.
\end{Definition}

\begin{figure*}[t]
    \centering
    \includegraphics[width=\textwidth]{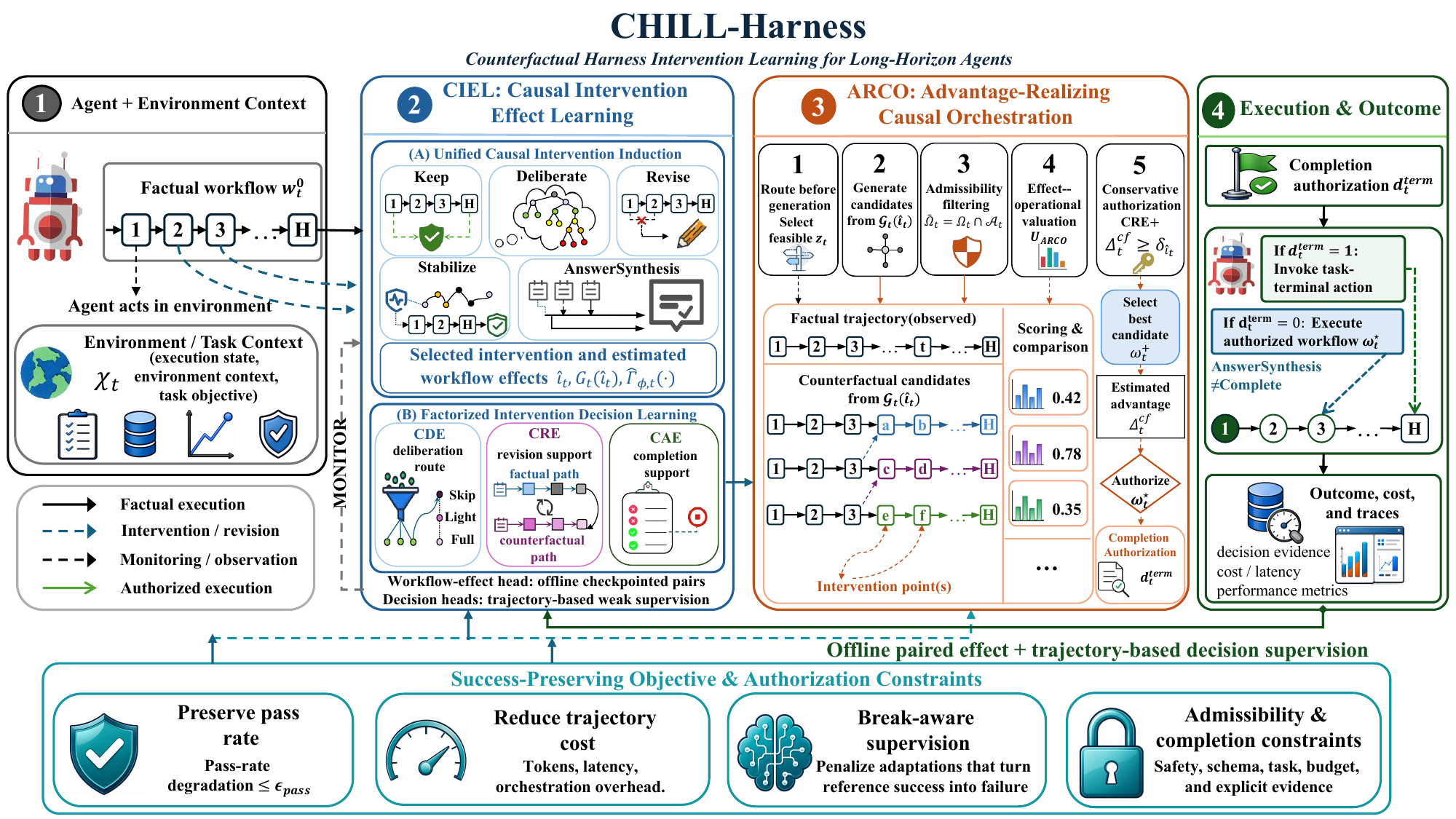}
    \caption{
    Overview of CHILL-Harness.
    Given context \(\chi_t\) and factual workflow \(\omega_t^0\), CIEL estimates
    workflow effects and predicts intervention, deliberation, revision, and
    completion signals.
    ARCO realizes them through route-before-generation, admissibility filtering,
    causal--operational valuation, and conservative authorization.
    Offline checkpointed paired executions supervise workflow-effect estimation,
    while trajectory outcomes provide weak supervision for the factorized
    decision heads.
    }
    \label{fig:chill_overview}
\end{figure*}

Following the preliminaries, \(\chi_t\), \(\omega_t^0\), \(\mathcal A_t\),
and \(\Gamma_t(\omega)\) denote the execution context, factual workflow,
admissible workflow set, and expected gain of \(\omega\) over
\(\omega_t^0\), respectively.
As illustrated in Figure~\ref{fig:chill_overview}, CHILL-Harness addresses
the two problems through \textbf{Causal Intervention Effect Learning (CIEL)}
and \textbf{Advantage-Realizing Causal Orchestration (ARCO)}, under explicit
success-preserving workflow authorization constraints.

\subsection{CIEL: Causal Intervention Effect Learning}
\label{sec:unified_intervention}

CIEL learns workflow-grounded intervention preferences for the Harness
Intervention Effect Problem and converts them into factorized deliberation,
revision, and completion signals.

\paragraph{Unified Causal Intervention Induction.}
CIEL abstracts recurring harness adaptations, including additional reasoning,
workflow correction, execution recovery, and answer construction, into a
unified intervention space
\citep{wang2024agent,yang2024sweagent,wang2025oscar}.
Let
\begin{equation*}
\resizebox{\columnwidth}{!}{$
\begin{aligned}
\mathcal R
&=
\{
\textsc{Inspect},
\textsc{Verify},
\textsc{Visual}
\},
&
\mathcal Q
&=
\{
\textsc{Dedup},
\textsc{NoOpRecovery}
\}
\end{aligned}
$}
\end{equation*}
denote the revision and stabilization modes.
The intervention space and its workflow realization are
\begin{equation*}
\begin{aligned}
\mathcal I_t^{\mathrm{wf}}
={}&
\{
\textsc{Keep},
\textsc{Deliberate},
\textsc{AnswerSynthesis}
\}
\\
&\cup
\{
\textsc{Revise}(r)
\mid
r\in\mathcal R
\}
\cup
\{
\textsc{Stabilize}(q)
\mid
q\in\mathcal Q
\},
\\
\mathcal G_t(\iota)
\equiv{}&
\mathcal G
\left(
\iota;\chi_t,\omega_t^0
\right)
\subseteq
\mathcal A_t,
\qquad
\mathcal G_t(\textsc{Keep})
=
\{\omega_t^0\}.
\end{aligned}
\end{equation*}
Here, \(\mathcal I_t^{\mathrm{wf}}\) specifies adaptation intent and
\(\mathcal G_t(\iota)\) maps it to admissible workflows.
\textsc{Keep} preserves the factual workflow;
\textsc{Deliberate} requests additional reasoning;
\textsc{Revise} acquires, verifies, or visually inspects evidence;
\textsc{Stabilize} suppresses repetition or recovers ineffective execution;
and \textsc{AnswerSynthesis} redirects execution toward nonterminal answer construction.
Each intervention inherits its causal contribution from its realized
workflow (details in Appendix A.2).

CIEL estimates the context-conditioned workflow effect as
\(
\widehat\Gamma_\phi
(
\omega;\chi_t,\omega_t^0
)
\approx
\Gamma_t(\omega)
\),
with the factual workflow serving as the zero-effect reference.
The estimates induce intervention-family preferences that are amortized into
online prediction:
\begin{subequations}
\begin{align}
\widehat V_t^{\mathrm{int}}(\iota)
&=
\max_{\omega\in\mathcal G_t(\iota)}
\widehat\Gamma_\phi
\left(
\omega;\chi_t,\omega_t^0
\right),
\label{eq:intervention_family_score}
\\
\iota_t^*
&=
\arg\max_{\iota\in\mathcal I_t^{\mathrm{wf}}}
\widehat V_t^{\mathrm{int}}(\iota),
\label{eq:offline_intervention_target}
\\
\widehat\iota_t
&=
\arg\max_{\iota\in\mathcal I_t^{\mathrm{wf}}}
p_{\theta_u}
\left(
\iota
\mid
\chi_t,\omega_t^0
\right).
\label{eq:intervention_prediction}
\end{align}
\end{subequations}
Equations~\eqref{eq:intervention_family_score}--\eqref{eq:intervention_prediction}
separate workflow-effect aggregation, offline intervention induction, and
amortized online prediction.
Because \(\mathcal G_t(\textsc{Keep})=\{\omega_t^0\}\), conservative
tie-breaking selects \textsc{Keep} whenever no alternative family has a
positive estimated effect. The workflow-effect estimator learns from checkpointed paired records, while
the intervention predictor amortizes the induced offline preference:
\begin{align*}
\mathcal L_{\mathrm{eff}}
&=
\mathbb E_{\mathcal D_{\mathrm{pair}}}
\left[
w_{t,\omega}
\left(
\widehat\Gamma_\phi
\left(
\omega;\chi_t,\omega_t^0
\right)
-
\widetilde\Gamma_t^{\mathrm{pair}}(\omega)
\right)^2
\right],
\\
\mathcal L_{\mathrm{int}}
&=
-
\log
p_{\theta_u}
\left(
\iota_t^*
\mid
\chi_t,\omega_t^0
\right).
\end{align*}
Here, \(\mathcal D_{\mathrm{pair}}\) contains factual--candidate executions
restored from the same context,
\(\widetilde\Gamma_t^{\mathrm{pair}}(\omega)\) is their empirical utility
difference, and \(w_{t,\omega}\) is its reliability weight.

For analysis, define
$
V_t^{\mathrm{int}}(\iota)
=
\max_{\omega\in\mathcal G_t(\iota)}
\Gamma_t(\omega),
$
$
\iota_t^\dagger
=
\arg\max_{\iota\in\mathcal I_t^{\mathrm{wf}}}
V_t^{\mathrm{int}}(\iota),
$
and let $
\mathcal C_t
=
\bigcup_{\iota\in\mathcal I_t^{\mathrm{wf}}}
\mathcal G_t(\iota).
$
The workflow-effect error and amortization gap are
\begin{equation}
\resizebox{\columnwidth}{!}{$
\begin{aligned}
\varepsilon_t
&=
\sup_{\omega\in\mathcal C_t}
\left|
\widehat\Gamma_\phi
\left(
\omega;\chi_t,\omega_t^0
\right)
-
\Gamma_t(\omega)
\right|,
&
\rho_t
&=
\widehat V_t^{\mathrm{int}}(\iota_t^*)
-
\widehat V_t^{\mathrm{int}}(\widehat\iota_t).
\end{aligned}
$}
\end{equation}
Then
\begin{equation}
\label{eq:amortized_intervention_result}
V_t^{\mathrm{int}}(\iota_t^\dagger)
-
V_t^{\mathrm{int}}(\widehat\iota_t)
\leq
2\varepsilon_t+\rho_t.
\end{equation}
Thus, effect-estimation and amortization errors jointly control intervention
selection regret; Appendix A.3 proves the result, and Appendix A.4 details paired-effect construction.

\paragraph{Factorized Intervention Decision Learning.}
CIEL factorizes online control into three decisions governing additional
deliberation, workflow revision, and task completion.

\textbf{Counterfactual Deliberation Effect (CDE).}
CDE addresses harness over-deliberation, where costly planning is repeatedly
invoked even when the factual workflow is already sufficient.
It predicts the required deliberation route:
\begin{equation}
\label{eq:cde_prediction}
p_{\theta_p}
\left(
z_t
\mid
\chi_t,\omega_t^0,\widehat\iota_t
\right),
\qquad
z_t
\in
\mathcal Z
=
\{
\textsc{Skip},
\textsc{Light},
\textsc{Full}
\}.
\end{equation}
\textsc{Skip} bypasses candidate generation,
\textsc{Light} enables restricted diagnostics, and
\textsc{Full} enables broader evaluation.
Thus, the intervention family specifies \emph{what} adaptation is useful,
whereas CDE determines \emph{how much} computation realizes it.

\textbf{Counterfactual Revision Effect (CRE).}
CRE addresses both workflow inertia and harmful replacement: the factual
workflow may require correction, while an unsupported alternative may further
degrade execution.
Given an alternative workflow \(\omega\), CRE predicts whether it provides
sufficient support for replacing the factual workflow:
\begin{equation}
\label{eq:cre_prediction}
p_{\theta_c}
\left(
y_t^{\mathrm{chg}}
\mid
\chi_t,\omega_t^0,\omega,\widehat\iota_t
\right),
\qquad
y_t^{\mathrm{chg}}
\in
\{
\textsc{Keep},
\textsc{Change}
\}.
\end{equation}
\textsc{Keep} retains \(\omega_t^0\), whereas \textsc{Change} supports
replacement by the evaluated alternative.

\textbf{Completion Attribution Effect (CAE).}
CAE predicts completion support to prevent both continued execution after task success and termination triggered solely by futility:
\begin{equation}
\label{eq:cae_prediction}
p_{\theta_e}
\left(
b_t
\mid
\chi_t,o_t,h_t
\right),
\qquad
b_t\in\{0,1\}.
\end{equation}
Here, \(o_t\) and \(h_t\) denote the latest observation and recent execution
summary; \(b_t=1\) supports completion, which still requires explicit task
evidence.

Logged execution traces provide weak targets
\(z_t^*\), \(y_t^{\mathrm{chg},*}\), and \(b_t^*\) for CDE, CRE, and CAE,
respectively.
Here, \(z_t^*\) denotes the preferred deliberation route,
\(y_t^{\mathrm{chg},*}\) denotes the preferred revision decision, and
\(b_t^*\) denotes whether retrospective task evidence supports completion.
Efficiency-related decisions receive positive supervision only when task
success is preserved.
The corresponding losses are
\[
\begin{aligned}
\mathcal L_{\mathrm{CDE}}
&=
-
\log
p_{\theta_p}
\left(
z_t^*
\mid
\chi_t,\omega_t^0,\widehat\iota_t
\right),
\\
\mathcal L_{\mathrm{CRE}}
&=
-
\log
p_{\theta_c}
\left(
y_t^{\mathrm{chg},*}
\mid
\chi_t,\omega_t^0,\omega_t^+,\widehat\iota_t
\right),
\\
\mathcal L_{\mathrm{CAE}}
&=
-
\log
p_{\theta_e}
\left(
b_t^*
\mid
\chi_t,o_t,h_t
\right).
\end{aligned}
\]
Thus, \(\mathcal L_{\mathrm{CDE}}\) supervises the amount of additional
reasoning, \(\mathcal L_{\mathrm{CRE}}\) supervises factual-versus-candidate
workflow selection, and \(\mathcal L_{\mathrm{CAE}}\) supervises
evidence-attributed completion.
The complete CIEL objective is
\begin{equation}
\label{eq:ciel_objective}
\mathcal L_{\mathrm{CIEL}}
=
\mathcal L_{\mathrm{eff}}
+
\mathcal L_{\mathrm{int}}
+
\mathcal L_{\mathrm{CDE}}
+
\mathcal L_{\mathrm{CRE}}
+
\mathcal L_{\mathrm{CAE}}.
\end{equation}
Equation~\eqref{eq:ciel_objective} jointly supervises workflow effects,
intervention preferences, deliberation, revision, and completion.
Detailed weak-target construction is provided in Appendix A.4.

\paragraph{Remark.}
CIEL separates \emph{what} adaptation is beneficial from \emph{how} it should be realized: workflow-effect learning induces intervention preferences, while CDE, CRE, and CAE factorize deliberation, revision, and completion decisions.

\subsection{ARCO: Advantage-Realizing Causal Orchestration}
\label{sec:orchestration}

ARCO addresses the Harness Intervention Realization Problem through route
selection, candidate valuation, conservative revision, and
evidence-grounded completion.
Let
\(
\widehat\Gamma_{\phi,t}(\omega)
:=
\widehat\Gamma_\phi
(
\omega;\chi_t,\omega_t^0
)
\).
Then
\[
\operatorname{ARCO}
\left(
\chi_t,
\omega_t^0,
o_t,
h_t,
\widehat\iota_t,
\widehat\Gamma_{\phi,t}(\cdot)
\right)
\longmapsto
\left(
\omega_t^\star,
d_t^{\mathrm{term}}
\right).
\]

Using the CDE predictor defined in Equation~\eqref{eq:cde_prediction}, ARCO
first selects the provisional deliberation route:
\begin{equation*}
\widetilde z_t
=
\arg\max_{z\in\mathcal Z}
p_{\theta_p}
\left(
z
\mid
\chi_t,\omega_t^0,\widehat\iota_t
\right).
\end{equation*}
Under
\(
\textsc{Skip}\prec\textsc{Light}\prec\textsc{Full},
\)
\(z_t\) is the highest feasible route not exceeding
\(\widetilde z_t\).
\textsc{Skip} retains \(\omega_t^0\); otherwise, the selected route exposes
$
\Omega_t
\subseteq
\{\omega_t^0\}
\cup
\mathcal G_t(\widehat\iota_t),
\overline\Omega_t
=
\Omega_t\cap\mathcal A_t,
$
where \(\overline\Omega_t\) contains the admissible route-exposed workflows.
Route feasibility is detailed in Appendix B.1.

Each admissible candidate is valued by combining operational utility and estimated workflow effect:
\begin{equation}
\label{eq:arco_utility}
\widehat U_{\mathrm{ARCO}}
\left(
\omega
\mid
\chi_t,z_t,\widehat\Gamma_{\phi,t}
\right)
=
\widehat U_{\mathrm{op}}
\left(
\omega\mid\chi_t,z_t
\right)
+
\eta_\Gamma
\widehat\Gamma_{\phi,t}(\omega).
\end{equation}
Here, \(\widehat U_{\mathrm{op}}\) aggregates progress, cost, risk,
information gain, robustness, and safety, while \(\eta_\Gamma\geq0\) weights estimated intervention evidence; Appendix B.2 gives the full construction.

ARCO selects the highest-valued workflow among the admissible route-exposed
candidates:
\begin{equation*}
\omega_t^+
=
\arg\max_{\omega\in\overline\Omega_t}
\widehat U_{\mathrm{ARCO}}
\left(
\omega
\mid
\chi_t,z_t,\widehat\Gamma_{\phi,t}
\right),
\end{equation*}
with advantage over the factual workflow
\begin{equation*}
\resizebox{\columnwidth}{!}{$
\begin{aligned}
\Delta_t^{\mathrm{cf}}
=
\widehat U_{\mathrm{ARCO}}
\left(
\omega_t^+
\mid
\chi_t,z_t,\widehat\Gamma_{\phi,t}
\right)
-
\widehat U_{\mathrm{ARCO}}
\left(
\omega_t^0
\mid
\chi_t,z_t,\widehat\Gamma_{\phi,t}
\right).
\end{aligned}
$}
\end{equation*}

Instantiating the CRE predictor in Equation~\eqref{eq:cre_prediction} with
the selected candidate \(\omega_t^+\), ARCO obtains
\[
y_t^{\mathrm{chg}}
=
\arg\max_{y\in\{\textsc{Keep},\textsc{Change}\}}
p_{\theta_c}
\left(
y
\mid
\chi_t,\omega_t^0,\omega_t^+,\widehat\iota_t
\right).
\]
The candidate is executed only when both CRE and its estimated advantage
support revision:
\begin{equation}
\label{eq:arco_revision}
\omega_t^\star
=
\begin{cases}
\omega_t^+,
&
y_t^{\mathrm{chg}}=\textsc{Change}
\ \land\
\Delta_t^{\mathrm{cf}}
\geq
\delta_{\widehat\iota_t},
\\[1mm]
\omega_t^0,
&
\text{otherwise}.
\end{cases}
\end{equation}
In Equation~\eqref{eq:arco_revision},
\(\delta_{\widehat\iota_t}\geq0\) is an intervention-dependent authorization
margin.
ARCO executes \(\omega_t^+\) only when both the CRE decision and its
estimated advantage support replacement; otherwise, it retains
\(\omega_t^0\) (see Appendix B.3 for details).

Finally, ARCO instantiates the CAE predictor in
Equation~\eqref{eq:cae_prediction} and combines its output with explicit task
evidence:
\begin{equation*}
\begin{aligned}
\widehat b_t
&=
\arg\max_{b\in\{0,1\}}
p_{\theta_e}
\left(
b
\mid
\chi_t,o_t,h_t
\right),
\\
e_t^{\mathrm{comp}}
&=
\mathbb I
\left[
\mathcal V_t
\left(
\chi_t,o_t,h_t
\right)
=1
\right],
\\
d_t^{\mathrm{term}}
&=
\widehat b_t e_t^{\mathrm{comp}}.
\end{aligned}
\end{equation*}
Here, \(\mathcal V_t\) is the task verifier and
\(\mathbb I[\cdot]\) is the binary indicator.

\paragraph{Remark.}
ARCO realizes learned effects conservatively: routing limits
candidate-generation cost, causal--operational valuation ranks admissible
alternatives, and advantage-margin authorization retains the factual workflow
when evidence is insufficient.

\subsection{Success-Preserving Objective and Authorization Constraints}
\label{sec:efficiency_safe}

Let \(\tau\) be a complete execution trajectory,
\(\pi_{\mathrm{CHILL},\Theta}\) the policy induced by CIEL and ARCO,
\(\Theta=\{\phi,\theta_u,\theta_p,\theta_c,\theta_e\}\), and
\(\pi_{\mathrm{ref}}\) the matched reference harness.
CHILL-Harness minimizes trajectory cost subject to bounded success degradation:
\begin{equation*}
\begin{aligned}
\min_{\Theta}\quad&
\mathbb E_{\tau\sim\pi_{\mathrm{CHILL},\Theta}}
\left[
\mathcal J_{\mathrm{eff}}(\tau)
\right]
\\
\text{s.t.}\quad&
\operatorname{PassRate}
\left(
\pi_{\mathrm{CHILL},\Theta}
\right)
\geq
\operatorname{PassRate}
\left(
\pi_{\mathrm{ref}}
\right)
-\epsilon_{\mathrm{pass}},
\end{aligned}
\end{equation*}
where \(\mathcal J_{\mathrm{eff}}(\tau)\) aggregates trajectory-level
resource cost and \(\epsilon_{\mathrm{pass}}\geq0\) is the allowed pass-rate degradation.

At each step, only admissible route-exposed workflows may be executed, and
termination requires both learned and verified completion evidence:
\begin{equation*}
\omega_t^\star
\in
\overline\Omega_t
=
\Omega_t\cap\mathcal A_t,
\qquad
d_t^{\mathrm{term}}
=
\widehat b_t e_t^{\mathrm{comp}}.
\end{equation*}

Because environment execution is not directly differentiable, the constrained
objective is realized through outcome-conditioned weak supervision rather
than trajectory-level backpropagation.
Let \(Y_\tau,Y_{\mathrm{ref}}\in\{0,1\}\) denote the current and reference
trajectory outcomes, and let
\(\lambda_{\mathrm{fail}},\lambda_{\mathrm{break}}\geq0\) be their penalty
weights. We define
\begin{equation}
\label{eq:trajectory_supervision}
\resizebox{\columnwidth}{!}{$
\begin{aligned}
\mathcal L_{\mathrm{traj}}(\tau)
=
\mathcal J_{\mathrm{eff}}(\tau)
+
\lambda_{\mathrm{fail}}
\left(
1-Y_\tau
\right)
+
\lambda_{\mathrm{break}}
\mathbb I
\left[
Y_{\mathrm{ref}}=1
\land
Y_\tau=0
\right].
\end{aligned}
$}
\end{equation}
Equation~\eqref{eq:trajectory_supervision} penalizes resource cost, task failure, and success-breaking adaptation, and constructs weak targets for CDE, CRE, and CAE.

\paragraph{Remark.}
The success-preserving objective and authorization constraints jointly protect task performance: efficiency-related decisions receive positive supervision only when success is preserved, while unsupported workflow revision and premature termination are disallowed.

Algorithm~\ref{alg:chill} summarizes the inference flow; Appendices
C.1--C.4 provide the complete objectives and authorizations.

\begin{algorithm}[t]
\caption{CHILL-Harness Overview (Full Procedure in Appendix C.4)}
\label{alg:chill}
\begin{algorithmic}[1]
\STATE \textbf{Input}: \(\chi_t,\omega_t^0,o_t,h_t\);
\textbf{Output}: \(\omega_t^\star,d_t^{\mathrm{term}}\)

\STATE
\(
(\widehat\iota_t,\widetilde z_t)
\leftarrow
\operatorname{CIEL}(\chi_t,\omega_t^0)
\)

\STATE Set \(z_t\) to the highest feasible route not exceeding
\(\widetilde z_t\)

\IF{\(z_t=\textsc{Skip}\)}
    \STATE \(\Omega_t\leftarrow\{\omega_t^0\}\)
\ELSE
    \STATE Generate \(\Omega_t\) from
    \(\mathcal G_t(\widehat\iota_t)\cup\{\omega_t^0\}\)
\ENDIF

\STATE
\(
\overline\Omega_t\leftarrow\Omega_t\cap\mathcal A_t
\);
select the highest-valued candidate \(\omega_t^+\)

\STATE Authorize \(\omega_t^\star\) using CRE and candidate advantage

\STATE Authorize \(d_t^{\mathrm{term}}\) using CAE and completion evidence

\IF{\(d_t^{\mathrm{term}}=1\)}
    \STATE Invoke the task-terminal action
\ELSE
    \STATE Execute \(\omega_t^\star\)
\ENDIF

\STATE Log execution evidence
\end{algorithmic}
\end{algorithm}

\begin{table*}[t]
\centering
\small
\setlength{\tabcolsep}{4pt}

\begin{threeparttable}
\begin{tabular}{@{}l l c c c c c c@{}}
\toprule
Benchmark
& Method
& Success
& Tokens
& \makecell[c]{Token Gain\\vs. Min.-Token}
& Tokens/Solved
& Rel. Runtime
& \makecell[c]{Time Gain\\vs. Min.-Runtime} \\
\midrule

\multirow{4}{*}{\textbf{GAIA}}
& Terminus-KIRA
& 70.2\%
& 134M
& --
& 2.03M
& $1.00\times$
& -- \\

& AWorld
& 69.7\%
& 161M
& --
& 2.45M
& $1.25\times$
& -- \\

& OWL Workforce
& 69.1\%
& 174M
& --
& 2.68M
& $1.30\times$
& -- \\

& \textbf{CHILL-Harness}
& \textbf{71.3\%}
& \textbf{96M}
& \textbf{28.4\%$\downarrow$}
& \textbf{1.43M}
& \textbf{$0.534\times$}
& \textbf{46.6\%$\downarrow$} \\

\midrule

\multirow{4}{*}{\makecell[l]{\textbf{SWE-bench}\\\textbf{Verified}}}
& Terminus-KIRA
& 65.0\%
& 157M
& --
& 2.42M
& $1.00\times$
& -- \\

& OpenHands
& 65.4\%
& 165M
& --
& 2.52M
& $1.10\times$
& -- \\

& CodeSweep--SWE-agent
& 53.4\%
& 153M
& --
& 2.87M
& $0.92\times$
& -- \\

& \textbf{CHILL-Harness}
& \textbf{65.6\%}
& \textbf{133M}
& \textbf{13.1\%$\downarrow$}
& \textbf{2.25M}
& \textbf{$0.682\times$}
& \textbf{25.9\%$\downarrow$} \\

\midrule

\multirow{4}{*}{\makecell[l]{\textbf{Terminal-}\\\textbf{Bench 2.0}}}
& Terminus-KIRA
& 49.4\%
& 291M
& --
& 7.11M
& $1.00\times$
& -- \\

& Meta-Harness
& 52.6\%
& 340M
& --
& 5.00M
& $1.25\times$
& -- \\

& LemonHarness
& 57.4\%
& 380M
& --
& 5.05M
& $1.40\times$
& -- \\

& \textbf{CHILL-Harness}
& \textbf{56.1\%}
& \textbf{224M}
& \textbf{23.1\%$\downarrow$}
& \textbf{4.87M}
& \textbf{$0.974\times$}
& \textbf{2.6\%$\downarrow$} \\

\bottomrule
\end{tabular}

\caption{
Effectiveness and efficiency across three long-horizon benchmarks.
Token Gain and Time Gain use the eligible non-CHILL baseline with the lowest
token consumption and runtime, respectively.
Relative runtime is normalized to Terminus-KIRA within each benchmark.
Token totals are rounded to the nearest million, whereas Tokens/Solved is
reported to two decimal places.
CHILL-Harness token totals include all additional model calls for
counterfactual deliberation and candidate valuation.}
\label{tab:chill_vs_baselines_three_benchmarks}
\end{threeparttable}
\end{table*}

\section{Experiments}
\label{sec:experiments}

We evaluate whether CHILL-Harness resolves the causal harness orchestration problem by learning beneficial workflow interventions, realizing them efficiently and admissibly, and preserving task success.
We examine these requirements through three empirical questions:

\begin{itemize}
    \item \textbf{RQ1: Effectiveness.}
    Does solving the Harness Intervention Effect Problem in
    Definition~\ref{def:harness_intervention_effect} identify adaptations
    that preserve or improve task success?

    \item \textbf{RQ2: Efficiency.}
    Does solving the Harness Intervention Realization Problem in
    Definition~\ref{def:harness_intervention_realization} reduce reasoning
    and execution cost without premature failure or termination?

    \item \textbf{RQ3: Generalization.}
    Do the effectiveness and efficiency benefits hold across heterogeneous
    execution environments?
\end{itemize}

We conduct baseline comparisons to verify that CHILL-Harness preserves
effectiveness while improving efficiency across benchmarks, and perform an
\textsc{Always-Full} ablation to establish the contribution of
counterfactual orchestration.

\subsection{Experimental Setup}
\label{sec:experimental_setup}

\paragraph{Benchmarks.}
We evaluate CHILL-Harness on
\textbf{GAIA} for information seeking and tool-assisted reasoning
\citep{mialon2023gaia},
\textbf{SWE-bench Verified} for repository-level software repair
\citep{jimenez2024swebench,openai2024swebenchverified}, and
\textbf{Terminal-Bench 2.0} for long-horizon terminal interaction
\citep{merrill2026terminalbench}.
These benchmarks cover deliberation, revision, stabilization, verification,
and completion under distinct execution environments.

\paragraph{Baselines.}
Each CHILL-Harness evaluation is paired with a reference run using the same model, tools, environment, task, and resource limits.
For external comparison, we report Terminus-KIRA on all benchmarks
\citep{terminuskira2026};
AWorld and OWL Workforce on GAIA
\citep{yu2025aworldorchestratingtrainingrecipe,hu2025owl};
OpenHands and CodeSweep--SWE-agent on SWE-bench Verified
\citep{wang2024openhands,yang2024sweagent,kimiteam2025kimik2}; and
Meta-Harness and LemonHarness on Terminal-Bench 2.0
\citep{lee2026metaharnessendtoendoptimizationmodel,
ren2026lemonharnesstechnicalreport}.

\begin{figure}[t]
    \centering
    \includegraphics[width=0.5\textwidth]
    {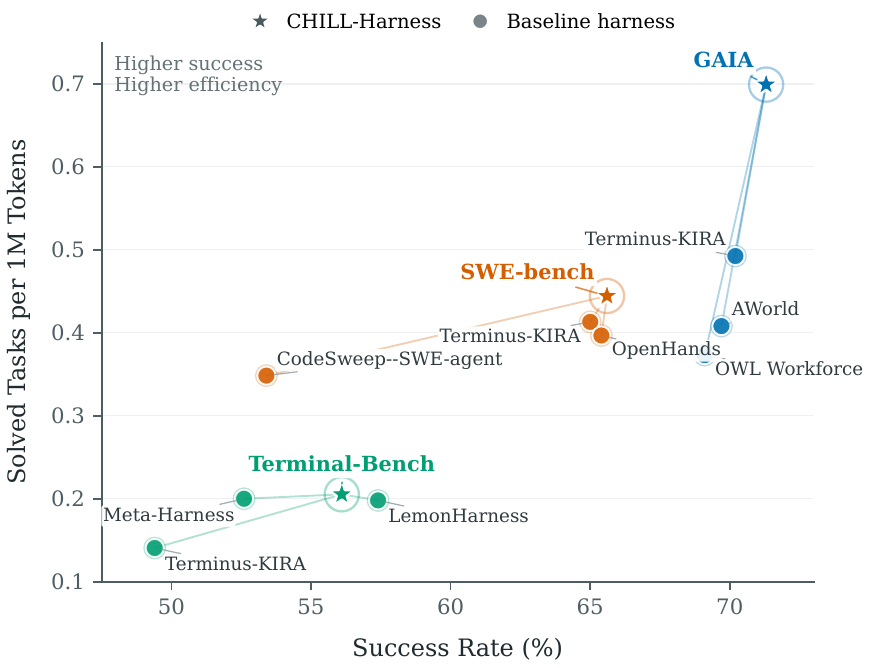}
    \caption{
    Joint effectiveness--efficiency comparison on GAIA, SWE-bench Verified,
    and Terminal-Bench 2.0.
    The horizontal axis reports task success and the vertical axis reports
    solved tasks per one million tokens; higher values on both axes are
    preferred.
    Stars denote CHILL-Harness, circles denote public baseline harnesses, and colors group methods evaluated on the same benchmark.
    }
    \label{fig:chill_efficiency}
\end{figure}

\paragraph{Model Configuration.} 
GAIA uses deepseek-v4-flash, whereas SWE-bench Verified and
Terminal-Bench 2.0 use deepseek-v4-pro
\cite{deepseekai2026deepseekv4}.
CIEL is trained offline on disjoint paired traces and frozen for evaluation;
checkpointed replay is disabled at test time.
All model calls use temperature \(1.0\) and at most \(150\) interaction turns.

\paragraph{Evaluation Metrics.}
Let \(m\) denote a method, \(N\) the number of tasks, and
\(y_i^{(m)}\in\{0,1\}\) whether \(m\) solves task \(i\).
Effectiveness is measured by
$
\operatorname{Success}(m)
=
\frac{1}{N}
\sum_{i=1}^{N}
y_i^{(m)}
\times 100\%.
$
Reasoning cost is measured by
$
\operatorname{Tokens}(m)=
\sum_{i=1}^{N}
\sum_{k\in\mathcal K_m}
\left(
\operatorname{Tok}^{\mathrm{in}}_{i,k}
+
\operatorname{Tok}^{\mathrm{out}}_{i,k}
\right),
\operatorname{Tokens/Solved}(m)=
\frac{
\operatorname{Tokens}(m)
}{
\sum_{i=1}^{N}y_i^{(m)}
},
$
where \(\mathcal K_m\) includes all model-based harness components.
Let \(T_i^{(m)}\) be the end-to-end runtime of method \(m\) on task \(i\), including inference, orchestration, tools, interaction, and verification:
$
\operatorname{RelativeRuntime}(m;b_T)
=
\frac{
\sum_{i=1}^{N}T_i^{(m)}
}{
\sum_{i=1}^{N}T_i^{(b_T)}
}.
$
For \(X\in\{\operatorname{Tokens},T\}\), resource gain is
$
\operatorname{Gain}_{X}(m;b_X)
=
\left(
1-
\frac{X(m)}{X(b_X)}
\right)
\times 100\%.
$
Here, \(b_T\) and \(b_X\) denote the runtime and resource reference baselines,
respectively.

\begin{table*}[t]
\centering
\small
\setlength{\tabcolsep}{4pt}

\begin{threeparttable}
\begin{tabular}{@{}l l c c c c c@{}}
\toprule
Benchmark
& Variant
& Success
& Tokens
& Tokens/Solved
& \makecell[c]{Full Calls /\\Revision Rate}
& Always-Full Overhead \\
\midrule

\multirow{2}{*}{\textbf{GAIA}}
& \textbf{CHILL-Harness}
& \textbf{71.3\%}
& \textbf{96.00M}
& \textbf{1.43M}
& --
& -- \\

& Always-Full
& 27.7\%
& 114.33M
& 4.40M
& 2,615 / 76.1\%
& \makecell[c]{18.33M tokens\\37.73 h} \\

\midrule

\multirow{2}{*}{\makecell[l]{\textbf{SWE-bench}\\\textbf{Verified}}}
& \textbf{CHILL-Harness}
& \textbf{65.6\%}
& \textbf{133.00M}
& \textbf{2.25M}
& --
& -- \\

& Always-Full
& 53.3\%
& 160.40M
& 3.34M
& 3,708 / 3.5\%
& \makecell[c]{27.40M tokens\\70.95 h} \\

\midrule

\multirow{2}{*}{\makecell[l]{\textbf{Terminal-}\\\textbf{Bench 2.0}}}
& \textbf{CHILL-Harness}
& \textbf{56.1\%}
& \textbf{224.19M}
& \textbf{4.87M}
& --
& -- \\

& Always-Full
& 56.3\%
& 259.77M
& 5.65M
& 4,234 / 3.9\%
& \makecell[c]{35.58M tokens\\91.98 h} \\

\bottomrule
\end{tabular}

\caption{
Ablation Experiment Results.
Full Calls / Revision Rate reports the number of full CDE events and the
fraction of those events that authorize workflow revision.
Always-Full Overhead reports the additional planner token consumption and
cumulative planner latency incurred by Always-Full relative to
CHILL-Harness on the same benchmark.
}
\label{tab:ablation_always_full}
\end{threeparttable}
\end{table*}

\subsection{Main Results}

Table~\ref{tab:chill_vs_baselines_three_benchmarks} reports effectiveness
and resource measurements against baselines, while
Figure~\ref{fig:chill_efficiency} summarizes their joint success--token
efficiency.
Controlled matched-reference comparisons assess success preservation,
whereas comparisons with public systems assess external competitiveness.

\paragraph{RQ1: Effectiveness.}
Under matched model, tool, environment, task, and resource settings,
CHILL-Harness preserves or improves task success on all three benchmarks.
It achieves the highest success rate among the compared systems on GAIA (\(71.3\%\)) and SWE-bench Verified (\(65.6\%\)).
On Terminal-Bench 2.0, its \(56.1\%\) success exceeds Terminus-KIRA and
Meta-Harness and remains close to LemonHarness (\(57.4\%\)).
Thus, the efficiency gains do not arise from broadly sacrificing task
effectiveness.

\paragraph{RQ2: Efficiency.}
CHILL-Harness achieves the highest number of solved tasks per one million
tokens in each benchmark group.
Against the lowest-token public baselines, it reduces token consumption by
\(28.4\%\), \(13.1\%\), and \(23.1\%\) on GAIA, SWE-bench Verified, and
Terminal-Bench 2.0, respectively.
Against the fastest eligible baselines, it reduces runtime by
\(46.6\%\), \(25.9\%\), and \(2.6\%\).

\paragraph{RQ3: Generalization.}
CHILL-Harness improves both success and token efficiency over all displayed
baselines on GAIA and SWE-bench Verified.
On Terminal-Bench 2.0, it achieves the highest token efficiency, while
LemonHarness retains a small success advantage, placing both systems on the
effectiveness--efficiency frontier.
This pattern holds across information seeking, software repair, and terminal
interaction.

\paragraph{Remark.}
Controlled comparisons support success preservation, while public-system
comparisons show that CHILL-Harness consistently improves token efficiency
and remains competitive in task effectiveness across heterogeneous
long-horizon environments.

\subsection{Ablation Experiment}
\label{sec:ablation_always_full}

We evaluate the contribution of CDE and its route-before-generation
realization in ARCO.
The \textsc{Always-Full} variant disables adaptive deliberation by setting \textbf{$z_t=\textsc{Full}$} whenever feasible, while retaining CIEL intervention prediction, CRE, and CAE,
as well as ARCO candidate generation, admissibility filtering, valuation, and
authorization.
The comparison therefore isolates the effect of selecting deliberation depth
before candidate generation.

Table~\ref{tab:ablation_always_full} shows that forced full deliberation
causes over-intervention on GAIA, increases cost while reducing success on
SWE-bench Verified, and incurs substantial planner overhead with little
benefit on Terminal-Bench 2.0.
These results validate the contribution of CDE and its route-before-generation
realization in ARCO.

\paragraph{Remark.}
CDE acts before candidate generation: CRE can reject unsupported workflow
revisions only after the corresponding candidate-generation cost has been
incurred, whereas route-before-generation suppresses low-value deliberation
and avoids the associated planner overhead.

\section{Conclusion}
\label{sec:conclusion}

We introduced CHILL-Harness, a causal framework that learns whether and how the harness should intervene
in the choice between factual and alternative workflows.
The empirical results support affirmative answers to our three research questions: CIEL identifies
interventions that preserve or improve task effectiveness, ARCO realizes them
with lower reasoning and execution cost, and these benefits generalize across
heterogeneous long-horizon environments.
Experiments on information seeking, software repair, and terminal interaction,
together with the \textsc{Always-Full} ablation, support both the overall effectiveness of CHILL-Harness and the necessity of selective counterfactual
deliberation.
We hope this work encourages the community to move beyond fixed harness
engineering toward learned, causal, and success-preserving orchestration,
providing a foundation for more efficient and reliable long-horizon agents.
\clearpage
\bibliography{aaai2027}

\clearpage


\clearpage


\setcounter{equation}{0}
\renewcommand{\theequation}{A\arabic{equation}}

\section*{Appendix A: Problem Formalization and CIEL Foundations}

This appendix provides the formal definitions, theoretical grounding, and
training implementation of Causal Intervention Effect Learning (CIEL).
Appendix A.1 formalizes the two harness intervention problems.
Appendix A.2 specifies the structural connection between semantic
interventions and executable workflows.
Appendix A.3 derives an error-dependent bound for amortized intervention
selection.
Appendix A.4 establishes paired-replay identification and details the
separated supervision of workflow effects and factorized decisions.

\subsection*{A.1 Formalization of the Harness Intervention Problems}

At execution step \(t\), let
\[
\chi_t
=
\left(
s_t,
c_t^{\mathrm{env}},
g_t
\right)
\]
denote the fixed execution context, where \(s_t\) is the execution state,
\(c_t^{\mathrm{env}}\) is the environment context, and \(g_t\) is the
current task objective.

Let \(\omega_t^0\) denote the factual workflow proposed without additional
harness adaptation, and let \(\mathcal A_t\) denote the workflows admissible
under the current task, environment, safety, and resource requirements, with
\[
\omega_t^0\in\mathcal A_t.
\]
The executed workflow is represented by \(W_t\), and its task- and
resource-aware execution performance is represented by \(U_t\).

For any \(\omega\in\mathcal A_t\),
\[
\operatorname{do}(W_t=\omega)
\]
denotes executing workflow \(\omega\) while holding \(\chi_t\) fixed.
All potential outcomes are evaluated under the same continuation policy
\(\pi^{\mathrm{cont}}\), evaluation horizon, resource-accounting rule, and
task evaluator.
\(U_t(\omega)\) captures downstream task performance, whereas local cost,
risk, and execution-quality signals are represented separately by
\(\widehat U_{\mathrm{op}}\).

\paragraph{Definition A.1: Harness Intervention Effect Problem.}
For an admissible workflow
\[
\omega\in\mathcal A_t\setminus\{\omega_t^0\},
\]
its intervention effect relative to the factual workflow is
\begin{equation}
\label{eq:appendix_intervention_effect}
\resizebox{\columnwidth}{!}{$
\begin{aligned}
\Gamma_t(\omega)
=
\mathbb E
\left[
U_t
\mid
\operatorname{do}(W_t=\omega),
\chi_t
\right]
-
\mathbb E
\left[
U_t
\mid
\operatorname{do}(W_t=\omega_t^0),
\chi_t
\right].
\end{aligned}
$}
\end{equation}
A positive value indicates expected improvement relative to the factual
workflow, whereas a negative value indicates expected degradation.
The factual workflow is the zero-effect reference:
\[
\Gamma_t(\omega_t^0)=0.
\]
The estimand therefore assigns causal value to concrete executable workflows
rather than directly to abstract intervention labels.

\paragraph{Definition A.2: Harness Intervention Realization Problem.}
Let
\[
\widehat\Gamma_{\phi,t}(\omega)
:=
\widehat\Gamma_\phi
\left(
\omega;
\chi_t,
\omega_t^0
\right)
\]
denote the estimated workflow effect, and let
\(\overline\Omega_t\subseteq\mathcal A_t\) be the route-exposed admissible
workflow set with \(\omega_t^0\in\overline\Omega_t\).
The highest-valued exposed candidate is
\begin{equation}
\label{eq:appendix_realization_candidate}
\omega_t^+
=
\arg\max_{\omega\in\overline\Omega_t}
\widehat U_{\mathrm{ARCO}}
\left(
\omega
\mid
\chi_t,
z_t,
\widehat\Gamma_{\phi,t}
\right).
\end{equation}
The authorized workflow then satisfies
\begin{equation}
\label{eq:appendix_realization_output}
\omega_t^\star
\in
\left\{
\omega_t^0,
\omega_t^+
\right\}.
\end{equation}
Candidate selection and execution authorization are distinct:
selection identifies the strongest currently exposed candidate, whereas
replacement additionally requires admissibility, learned revision support,
and a sufficient estimated advantage.
The resulting orchestration chain is $
\text{workflow-effect estimation}
\longrightarrow
\text{route-exposed valuation}
\longrightarrow
\text{execution authorization}.
$

\subsection*{A.2 Workflow-Grounded Intervention Induction}

Let
\[
I_t\in\mathcal I_t^{\mathrm{wf}}
\]
denote the semantic harness intervention selected at step \(t\).
The intervention families have the following meanings.

\begin{itemize}
    \item \textsc{Keep} preserves the factual workflow \(\omega_t^0\).

    \item \textsc{Deliberate} allocates additional reasoning without
    prescribing a particular workflow revision.

    \item \textsc{Revise}(\textsc{Inspect}) acquires execution evidence
    missing from the current context.

    \item \textsc{Revise}(\textsc{Verify}) validates an intermediate state,
    tool result, or candidate answer.

    \item \textsc{Revise}(\textsc{Visual}) invokes visual processing when
    image or multimodal evidence is required.

    \item \textsc{Stabilize}(\textsc{Dedup}) suppresses repeated or
    semantically redundant operations.

    \item \textsc{Stabilize}(\textsc{NoOpRecovery}) redirects an execution
    that has produced an empty, ineffective, or stalled action.

    \item \textsc{AnswerSynthesis} redirects broad exploration toward
    evidence organization and answer construction without itself terminating
    execution.
\end{itemize}

These intervention families describe adaptation intents rather than unique
workflows.
The realization map
\[
\mathcal G_t(\iota)
=
\mathcal G
\left(
\iota;
\chi_t,
\omega_t^0
\right)
\subseteq
\mathcal A_t
\]
instantiates intervention \(\iota\) according to the current context,
factual workflow, available tools, and task constraints.
Unavailable intervention families are removed from the set considered at step
\(t\), so \(\mathcal G_t(\iota)\neq\emptyset\) for every family entering a
maximization.

\paragraph{Implementation correspondence.}
The intervention predictor produces a discrete intervention-family label.
A workflow-construction component then converts the label into one or more
structured workflow candidates.
For \textsc{Keep}, the constructed set contains only \(\omega_t^0\).
For \textsc{Deliberate}, the harness allocates an additional reasoning step
without forcing a tool or workflow change.
For the three \textsc{Revise} modes, candidate templates constrain the next
workflow toward evidence inspection, result verification, or visual
analysis.
For the two \textsc{Stabilize} modes, recent execution traces are checked for
duplicate actions, repeated tool arguments, empty results, or stalled
progress.
\textsc{AnswerSynthesis} generates a nonterminal workflow that organizes
available evidence into a candidate answer.
Each generated workflow is parsed into the same internal representation used
by the base harness before ARCO filtering and valuation.

\paragraph{Assumption A.1: Workflow Mediation.}
For fixed \(\chi_t\), each
\(\omega\in\mathcal G_t(\iota)\) is an executable realization of
intervention \(\iota\), intervention \(I_t\) affects \(U_t\) only through
the realized workflow \(W_t\), and intervention consistency holds.
This is a structural modeling restriction rather than an identification
result: semantic labels index workflow families but are not assigned
context-independent causal values.

Under this assumption,
\begin{equation}
\label{eq:appendix_grounding_equivalence}
\resizebox{\columnwidth}{!}{$
\begin{aligned}
\mathbb E
\left[
U_t
\mid
\operatorname{do}(I_t=\iota,W_t=\omega),
\chi_t
\right]
=
\mathbb E
\left[
U_t
\mid
\operatorname{do}(W_t=\omega),
\chi_t
\right].
\end{aligned}
$}
\end{equation}
Hence, the effect associated with intervention \(\iota\) is evaluated through
its concrete realization \(\omega\), relative to \(\omega_t^0\).
The same intervention family may therefore induce different effects across
contexts and workflow realizations.

\subsection*{A.3 Error-Bounded Amortized Intervention Selection}

For fixed \((\chi_t,\omega_t^0)\), define the true and estimated values of
intervention family \(\iota\) as
\begin{align}
V_t^{\mathrm{int}}(\iota)
&=
\max_{\omega\in\mathcal G_t(\iota)}
\Gamma_t(\omega),
\label{eq:appendix_true_intervention_value}
\\
\widehat V_t^{\mathrm{int}}(\iota)
&=
\max_{\omega\in\mathcal G_t(\iota)}
\widehat\Gamma_\phi
\left(
\omega;
\chi_t,
\omega_t^0
\right).
\label{eq:appendix_estimated_intervention_value}
\end{align}

Let
\[
\mathcal C_t
=
\bigcup_{\iota\in\mathcal I_t^{\mathrm{wf}}}
\mathcal G_t(\iota)
\]
denote all workflows considered at step \(t\), including the factual
workflow, and define
\[
\varepsilon_t
=
\sup_{\omega\in\mathcal C_t}
\left|
\widehat\Gamma_\phi
\left(
\omega;
\chi_t,
\omega_t^0
\right)
-
\Gamma_t(\omega)
\right|.
\]

Let
\[
\iota_t^\dagger
=
\arg\max_{\iota\in\mathcal I_t^{\mathrm{wf}}}
V_t^{\mathrm{int}}(\iota)
\]
be a true best intervention family,
\[
\iota_t^*
=
\arg\max_{\iota\in\mathcal I_t^{\mathrm{wf}}}
\widehat V_t^{\mathrm{int}}(\iota)
\]
be the offline estimated target, and define the amortization gap
\[
\rho_t
=
\widehat V_t^{\mathrm{int}}(\iota_t^*)
-
\widehat V_t^{\mathrm{int}}(\widehat\iota_t)
\geq0.
\]

\paragraph{Proposition A.1: Effect-and-Amortization Error Bound.}
For any estimated workflow-effect function and online intervention predictor,
\begin{equation}
\label{eq:appendix_amortized_consistency}
V_t^{\mathrm{int}}(\iota_t^\dagger)
-
V_t^{\mathrm{int}}(\widehat\iota_t)
\leq
2\varepsilon_t+\rho_t.
\end{equation}

\paragraph{Proof.}
For every intervention family \(\iota\), the maximum operator and the
definition of \(\varepsilon_t\) imply
\[
\left|
\widehat V_t^{\mathrm{int}}(\iota)
-
V_t^{\mathrm{int}}(\iota)
\right|
\leq
\varepsilon_t.
\]
Therefore,
\[
\begin{aligned}
V_t^{\mathrm{int}}(\iota_t^\dagger)
&\leq
\widehat V_t^{\mathrm{int}}(\iota_t^\dagger)
+
\varepsilon_t
\\
&\leq
\widehat V_t^{\mathrm{int}}(\iota_t^*)
+
\varepsilon_t
\\
&=
\widehat V_t^{\mathrm{int}}(\widehat\iota_t)
+
\rho_t
+
\varepsilon_t
\\
&\leq
V_t^{\mathrm{int}}(\widehat\iota_t)
+
2\varepsilon_t
+
\rho_t.
\end{aligned}
\]
Rearranging proves
Equation~\eqref{eq:appendix_amortized_consistency}.
\hfill\(\square\)

\paragraph{Corollary A.1: Exact Intervention Recovery.}
Suppose that \(\iota_t^\dagger\) is unique and define its true margin as
\[
\Delta_t^{\mathrm{int}}
=
V_t^{\mathrm{int}}(\iota_t^\dagger)
-
\max_{\iota\neq\iota_t^\dagger}
V_t^{\mathrm{int}}(\iota).
\]
If
\[
\Delta_t^{\mathrm{int}}
>
2\varepsilon_t+\rho_t,
\]
then
\[
\widehat\iota_t
=
\iota_t^\dagger.
\]
The result does not assume that estimated and true rankings are identical.
Instead, it quantifies how workflow-effect estimation error and amortized
prediction error jointly determine intervention-selection quality.
When the online predictor reproduces the offline target, \(\rho_t=0\), and
the selection regret is bounded by \(2\varepsilon_t\).

\subsection*{A.4 CIEL Training and Separated Supervision}

CIEL separates workflow-effect supervision from factorized decision
supervision.
The workflow-effect estimator is trained only from offline checkpointed paired
executions that compare factual and alternative workflows from the same
execution context.
Retrospective trajectory evidence provides weak supervision for deliberation,
revision, and completion decisions, but is not treated as an identified
workflow-effect observation.

\paragraph{Checkpointed paired intervention records.}
For a selected training state, the harness checkpoints the execution context
\(\chi_t\), including the agent context, recent execution memory,
environment state, tool state, intermediate artifacts, and current task
progress.
The checkpoint is restored to execute the factual workflow \(\omega_t^0\)
and an admissible counterfactual candidate \(\omega\) as two isolated
branches.
Both branches use the same continuation policy \(\pi^{\mathrm{cont}}\),
evaluation horizon, model and tool configuration, task evaluator, and
resource-accounting rule.
Branch-specific planning, model inference, tool interaction, and execution
costs are included in the resulting utility.
Post-checkpoint randomness is either matched across branches or independently
sampled from the same distribution and remains independent of workflow
assignment.

The analysis set \(\mathcal C_t\) in Appendix A.3 intentionally includes the
factual workflow.
For paired replay, define only the counterfactual subset
\begin{equation}
\label{eq:appendix_counterfactual_candidate_set}
\mathcal C_t^{\mathrm{cf}}
=
\mathcal C_t\setminus\{\omega_t^0\}.
\end{equation}
No replay count is required for \(\omega_t^0\), whose relative effect is
fixed to zero.
If \(\mathcal C_t^{\mathrm{cf}}=\emptyset\), no paired-effect record is
constructed at that state and \textsc{Keep} remains the only available
reference family.

For each \(\omega\in\mathcal C_t^{\mathrm{cf}}\), let
\(K_{t,\omega}\geq1\) be the number of valid paired replays, and let
\[
U_t^{(k)}(\omega)
\quad\text{and}\quad
U_t^{(k)}(\omega_t^0)
\]
denote the candidate and factual utilities in replay \(k\).
The paired workflow-effect target is
\begin{equation}
\label{eq:appendix_paired_effect_target}
\widetilde\Gamma_t^{\mathrm{pair}}(\omega)
=
\frac{1}{K_{t,\omega}}
\sum_{k=1}^{K_{t,\omega}}
\left[
U_t^{(k)}(\omega)
-
U_t^{(k)}(\omega_t^0)
\right].
\end{equation}
The factual reference is fixed by construction:
\[
\widetilde\Gamma_t^{\mathrm{pair}}(\omega_t^0)
=
\widehat\Gamma_\phi
\left(
\omega_t^0;
\chi_t,
\omega_t^0
\right)
=
0.
\]

\paragraph{Assumption A.2: State-Sufficient Paired Replay.}
The restored checkpoint contains the pre-intervention variables required to
reproduce the execution context relevant to post-intervention performance.
After restoration, the branches differ only in their assigned workflow,
follow the same continuation and evaluation protocol, do not interfere, and
use post-checkpoint randomness independent of workflow assignment.
The candidate set, replay count, and replay-inclusion criteria are fixed
before paired outcomes are observed; replay validity does not depend on the
sign or magnitude of the observed utility difference.

\paragraph{Proposition A.2: Identification and Concentration by Paired Replay.}
Under Assumption A.2,
\begin{equation}
\label{eq:appendix_paired_unbiasedness}
\mathbb E
\left[
\widetilde\Gamma_t^{\mathrm{pair}}(\omega)
\mid
\chi_t
\right]
=
\Gamma_t(\omega)
\end{equation}
for every \(\omega\in\mathcal C_t^{\mathrm{cf}}\).

Suppose additionally that
\[
U_t(\omega)\in[u_{\min},u_{\max}],
\qquad
R=u_{\max}-u_{\min}.
\]
For nonempty \(\mathcal C_t^{\mathrm{cf}}\), define
\[
M_t
=
\left|\mathcal C_t^{\mathrm{cf}}\right|,
\qquad
K_t^{\mathrm{pair}}
=
\min_{\omega\in\mathcal C_t^{\mathrm{cf}}}
K_{t,\omega}.
\]
If paired replays are independent conditional on \(\chi_t\), then, with
probability at least \(1-\zeta\),
\begin{equation}
\label{eq:appendix_paired_concentration}
\sup_{\omega\in\mathcal C_t^{\mathrm{cf}}}
\left|
\widetilde\Gamma_t^{\mathrm{pair}}(\omega)
-
\Gamma_t(\omega)
\right|
\leq
R
\sqrt{
\frac{
2\log(2M_t/\zeta)
}{
K_t^{\mathrm{pair}}
}
}.
\end{equation}

\paragraph{Proof.}
Because both branches begin from the same restored context and workflow
assignment is independent of post-checkpoint randomness, each branch follows
the potential-outcome distribution associated with its assigned workflow.
Linearity of expectation yields
Equation~\eqref{eq:appendix_paired_unbiasedness}.
For each candidate, the paired difference lies in \([-R,R]\).
Hoeffding's inequality bounds the empirical-mean deviation, and a union bound
over the \(M_t\) counterfactual candidates yields
Equation~\eqref{eq:appendix_paired_concentration}.
\hfill\(\square\)

\paragraph{Connection to intervention-selection error.}
Define the effect-model fitting error over paired targets as
\[
\alpha_t
=
\sup_{\omega\in\mathcal C_t^{\mathrm{cf}}}
\left|
\widehat\Gamma_\phi
\left(
\omega;
\chi_t,
\omega_t^0
\right)
-
\widetilde\Gamma_t^{\mathrm{pair}}(\omega)
\right|.
\]
Because the factual effect and its estimate are both zero, the error
\(\varepsilon_t\) defined over the full set \(\mathcal C_t\) equals the
supremum over \(\mathcal C_t^{\mathrm{cf}}\) whenever this subset is
nonempty.
Therefore, with probability at least \(1-\zeta\),
\[
\varepsilon_t
\leq
\alpha_t
+
R
\sqrt{
\frac{
2\log(2M_t/\zeta)
}{
K_t^{\mathrm{pair}}
}
}.
\]
Combining this inequality with Proposition A.1 gives
\begin{equation}
\label{eq:appendix_end_to_end_regret}
\resizebox{\columnwidth}{!}{$
\begin{aligned}
V_t^{\mathrm{int}}(\iota_t^\dagger)
-
V_t^{\mathrm{int}}(\widehat\iota_t)
\leq
2\alpha_t
+
2R
\sqrt{
\frac{
2\log(2M_t/\zeta)
}{
K_t^{\mathrm{pair}}
}
}
+
\rho_t.
\end{aligned}
$}
\end{equation}
Thus, intervention-selection quality is jointly controlled by workflow-effect
fitting error, finite paired-replay error, and amortized prediction error.

\paragraph{Paired workflow-effect training set.}
The workflow-effect training set is
\[
\mathcal D_{\mathrm{pair}}
=
\left\{
\left(
\chi_t,
\omega_t^0,
\omega,
\widetilde\Gamma_t^{\mathrm{pair}}(\omega),
w_{t,\omega}
\right)
:
\omega\in\mathcal C_t^{\mathrm{cf}}
\right\},
\]
where \(w_{t,\omega}\geq0\) is a reliability weight determined from
pre-specified replay-validity diagnostics, replay count, and comparison
stability, without using the sign of the observed effect.
Pairs are excluded when the checkpoint cannot be restored, the branches do
not share the same continuation protocol, or branch outcomes cannot be
isolated.
Only checkpointed paired executions are treated as workflow-effect
observations.
Unpaired trajectories, incomplete comparisons, and model-scored candidates
do not directly supervise \(\widehat\Gamma_\phi\) and are not interpreted as
identified causal effects.

The workflow-effect estimator is trained by
\begin{equation}
\label{eq:appendix_effect_loss}
\resizebox{\columnwidth}{!}{$
\begin{aligned}
\mathcal L_{\mathrm{eff}}
=
\mathbb E_{
\left(
\chi_t,
\omega_t^0,
\omega,
\widetilde\Gamma_t^{\mathrm{pair}}(\omega),
w_{t,\omega}
\right)
\sim
\mathcal D_{\mathrm{pair}}
}
\left[
w_{t,\omega}
\left(
\widehat\Gamma_\phi
\left(
\omega;
\chi_t,
\omega_t^0
\right)
-
\widetilde\Gamma_t^{\mathrm{pair}}(\omega)
\right)^2
\right].
\end{aligned}
$}
\end{equation}

\paragraph{Intervention target and objective.}
Candidate workflows are grouped by semantic intervention family.
The offline target is induced by the learned workflow-effect estimator:
\begin{equation}
\label{eq:appendix_intervention_target}
\iota_t^*
=
\arg\max_{\iota\in\mathcal I_t^{\mathrm{wf}}}
\max_{\omega\in\mathcal G_t(\iota)}
\widehat\Gamma_\phi
\left(
\omega;
\chi_t,
\omega_t^0
\right).
\end{equation}
When no alternative family has a positive estimated effect, conservative
tie-breaking assigns
\[
\iota_t^*=\textsc{Keep}.
\]
The amortized intervention predictor is trained by
\begin{equation}
\label{eq:appendix_intervention_loss}
\mathcal L_{\mathrm{int}}
=
-
\log
p_{\theta_u}
\left(
\iota_t^*
\mid
\chi_t,
\omega_t^0
\right).
\end{equation}

\paragraph{Trajectory evidence for factorized decisions.}
For each intervention step, the implementation extracts
\[
\mathcal E_t^{\mathrm{dec}}
=
\left(
Y_\tau,
Y_{\mathrm{ref}},
E_t^{\mathrm{err}},
C_t^{\mathrm{main}},
C_t^{\mathrm{plan}},
r_t,
c_t,
\Delta_t^{\mathrm{cf}},
\mathcal H_t^{\mathrm{exec}}
\right),
\]
where \(Y_\tau\) and \(Y_{\mathrm{ref}}\) are the current and reference task
outcomes; \(E_t^{\mathrm{err}}\) records execution failure;
\(C_t^{\mathrm{main}}\) and \(C_t^{\mathrm{plan}}\) record main-agent and
planner costs; \(r_t\) and \(c_t\) summarize route and candidate statistics;
and \(\mathcal H_t^{\mathrm{exec}}\) contains progress, repetition, failure,
and completion evidence.
A decision-target operator produces
\[
\Psi_{\mathrm{dec}}
\left(
\mathcal E_t^{\mathrm{dec}}
\right)
=
\left(
z_t^*,
y_t^{\mathrm{chg},*},
b_t^*
\right).
\]
These are outcome-conditioned weak decision labels, not observations of
\(\Gamma_t(\omega)\).

\paragraph{Counterfactual Deliberation Effect target.}
The target
\[
z_t^*
\in
\left\{
\textsc{Skip},
\textsc{Light},
\textsc{Full}
\right\}
\]
specifies the preferred amount of additional deliberation.
\textsc{Skip} is assigned when the factual workflow is sufficient or
additional planning produces no useful decision evidence.
\textsc{Light} is assigned when restricted diagnostics or a small candidate
set is sufficient.
\textsc{Full} is assigned only when broader generation or evaluation
materially improves intervention selection, recovers ineffective execution,
or prevents a high-risk factual action.
Among routes preserving decision quality and task success, the target selects
the least costly route.

\paragraph{Counterfactual Revision Effect target.}
The target
\[
y_t^{\mathrm{chg},*}
\in
\left\{
\textsc{Keep},
\textsc{Change}
\right\}
\]
specifies whether the selected candidate \(\omega_t^+\) should replace the
factual workflow.
\textsc{Change} requires evidence that the candidate improves the factual
execution while preserving success and satisfying authorization conditions.
When paired execution evidence is available, its observed workflow difference
provides the primary comparison.
Unpaired retrospective evidence may provide auxiliary decision supervision
but does not enter \(\mathcal D_{\mathrm{pair}}\).
Unsafe, invalid, task-incompatible, over-budget, duplicate, ineffective, or
success-breaking candidates receive \textsc{Keep} supervision.

\paragraph{Completion Attribution Effect target.}
The target
\[
b_t^*\in\{0,1\}
\]
indicates whether task completion is supported by evidence already available
at step \(t\).
The label \(b_t^*=1\) requires retrospective evidence that the task
requirements were already satisfied, such as a valid final answer, passing
tests, a required artifact, or an environment-provided success state.
Budget exhaustion, inactivity, repeated failure, predicted futility, or the
absence of another action does not produce a positive completion target.

\paragraph{Head-specific objectives.}
The factorized decision heads are trained by
\begin{align}
\mathcal L_{\mathrm{CDE}}
&=
-
\log
p_{\theta_p}
\left(
z_t^*
\mid
\chi_t,
\omega_t^0,
\widehat\iota_t
\right),
\label{eq:appendix_cde_loss}
\\
\mathcal L_{\mathrm{CRE}}
&=
-
\log
p_{\theta_c}
\left(
y_t^{\mathrm{chg},*}
\mid
\chi_t,
\omega_t^0,
\omega_t^+,
\widehat\iota_t
\right),
\label{eq:appendix_cre_loss}
\\
\mathcal L_{\mathrm{CAE}}
&=
-
\log
p_{\theta_e}
\left(
b_t^*
\mid
\chi_t,
o_t,
h_t
\right).
\label{eq:appendix_cae_loss}
\end{align}

The complete CIEL objective is
\begin{equation}
\label{eq:appendix_ciel_objective}
\mathcal L_{\mathrm{CIEL}}
=
\mathcal L_{\mathrm{eff}}
+
\mathcal L_{\mathrm{int}}
+
\mathcal L_{\mathrm{CDE}}
+
\mathcal L_{\mathrm{CRE}}
+
\mathcal L_{\mathrm{CAE}}.
\end{equation}
The objectives use distinct sources of supervision:
\(\mathcal L_{\mathrm{eff}}\) learns workflow effects from offline
checkpointed paired executions;
\(\mathcal L_{\mathrm{int}}\) amortizes the preference induced by the learned
workflow effects; and
\(\mathcal L_{\mathrm{CDE}}\), \(\mathcal L_{\mathrm{CRE}}\), and
\(\mathcal L_{\mathrm{CAE}}\) learn route, revision, and completion decisions
from success-preserving trajectory evidence.
This separation prevents heuristic or model-scored trajectory signals from
being interpreted as identified workflow-effect observations.


\setcounter{equation}{0}
\renewcommand{\theequation}{B\arabic{equation}}

\section*{Appendix B: ARCO Realization and Authorization Details}

This appendix explains how Advantage-Realizing Causal Orchestration (ARCO)
implements CIEL decisions through route selection, candidate generation,
causal--operational valuation, workflow authorization, and completion
verification.

\subsection*{B.1 Route Feasibility and Candidate Construction}

Let
\[
\mathcal Z
=
\left\{
\textsc{Skip},
\textsc{Light},
\textsc{Full}
\right\},
\qquad
\textsc{Skip}
\prec
\textsc{Light}
\prec
\textsc{Full}.
\]
The ordering represents increasing deliberation depth, candidate-generation
coverage, and execution cost.
The provisional route is
\begin{equation}
\label{eq:appendix_provisional_route}
\widetilde z_t
=
\arg\max_{z\in\mathcal Z}
p_{\theta_p}
\left(
z
\mid
\chi_t,
\omega_t^0,
\widehat\iota_t
\right).
\end{equation}

Let \(C^{\mathrm{plan}}(z)\) be the predicted planning cost of route \(z\),
\(B_t^{\mathrm{plan}}\) the remaining planning budget,
\(N_t^{\mathrm{full}}\) the number of previous full-deliberation calls,
\(B^{\mathrm{full}}\) their maximum allowed number,
\(t_{\mathrm{last}}^{\mathrm{full}}\) the most recent full-deliberation step,
\(\kappa\) the full-route cooldown,
\(\mathcal T(z)\) the tools or capabilities required by route \(z\), and
\(\mathcal T_t^{\mathrm{avail}}\) the currently available tools and
capabilities.
If no full route has previously been executed, set
\(t_{\mathrm{last}}^{\mathrm{full}}=-\infty\).

Define the common planning-and-tool feasibility indicator as
\begin{equation}
\label{eq:appendix_route_base_feasibility}
g_t(z)
=
\mathbb I
\left[
C^{\mathrm{plan}}(z)
\leq
B_t^{\mathrm{plan}}
\right]
\mathbb I
\left[
\mathcal T(z)
\subseteq
\mathcal T_t^{\mathrm{avail}}
\right].
\end{equation}
The route feasibility is
\begin{equation}
\label{eq:appendix_route_feasibility}
f_t(z)
=
\begin{cases}
1,
&
z=\textsc{Skip},
\\
g_t(z),
&
z=\textsc{Light},
\\
g_t(z)\,
\mathbb I
\left[
\begin{array}{c}
N_t^{\mathrm{full}}<B^{\mathrm{full}},\\[-1mm]
t-t_{\mathrm{last}}^{\mathrm{full}}\geq\kappa
\end{array}
\right],
&
z=\textsc{Full}.
\end{cases}
\end{equation}
The feasible route set is
\[
\mathcal F_t
=
\left\{
z\in\mathcal Z:
f_t(z)=1
\right\}.
\]
Because \textsc{Skip} is always feasible, \(\mathcal F_t\neq\emptyset\).
The executed route is
\begin{equation}
\label{eq:appendix_route_guard}
z_t
=
\max_{\prec}
\left\{
z\in\mathcal F_t:
z\preceq\widetilde z_t
\right\}.
\end{equation}
Thus, an infeasible prediction is downgraded to the highest feasible route and
is never upgraded beyond the learned prediction.

The raw route-specific candidate set is
\begin{equation}
\label{eq:appendix_route_candidate_set}
\Omega_t^{\mathrm{raw}}
=
\begin{cases}
\{\omega_t^0\},
&
z_t=\textsc{Skip},
\\[1mm]
\{\omega_t^0\}
\cup
\mathcal G_t^{\mathrm{light}}
\left(
\widehat\iota_t
\right),
&
z_t=\textsc{Light},
\\[1mm]
\{\omega_t^0\}
\cup
\mathcal G_t
\left(
\widehat\iota_t
\right),
&
z_t=\textsc{Full}.
\end{cases}
\end{equation}
Here,
\(\mathcal G_t^{\mathrm{light}}(\widehat\iota_t)\subseteq
\mathcal G_t(\widehat\iota_t)\)
is a restricted, low-cost realization set.
Candidate exposure is bounded by
\[
K_{\textsc{Skip}}=1,
\qquad
K_{\textsc{Light}}<K_{\textsc{Full}}.
\]
These route budgets are distinct from the paired-replay quantity
\(K_t^{\mathrm{pair}}\) in Appendix A.4.

Let
\(\widehat S_t(\omega\mid\chi_t)\) be the workflow-safety score and
\(\xi_t\) its minimum accepted threshold.
Before expensive valuation, the retained candidate set is
\begin{equation}
\label{eq:appendix_prescreened_candidates}
\resizebox{\columnwidth}{!}{$
\begin{aligned}
\Omega_t
=
\{\omega_t^0\}
\cup
\operatorname{TopK}_{K_{z_t}-1}
\left
\{
\omega
\in
\Omega_t^{\mathrm{raw}}\setminus\{\omega_t^0\}:
\widehat S_t(\omega\mid\chi_t)
\geq
\xi_t
\right\},
\end{aligned}
$}
\end{equation}
where \(\operatorname{TopK}_0(\cdot)=\emptyset\). When more candidates pass the safety threshold than the route budget allows,
\(\operatorname{TopK}\) retains them according to the low-cost
workflow-construction score produced before full valuation.
The admissible route-exposed set is
\begin{equation}
\label{eq:appendix_filtered_candidates}
\overline\Omega_t
=
\Omega_t
\cap
\mathcal A_t.
\end{equation}
Because \(\omega_t^0\in\mathcal A_t\), the filtered set is nonempty.

\paragraph{Implementation correspondence.}
The CDE head outputs scores for the three route classes.
A deterministic feasibility stage checks route-specific planning budgets,
full-route limits and cooldown, required tools, and environment capabilities.
Under \textsc{Skip}, planner and candidate-generation calls are bypassed.
Under \textsc{Light}, the implementation uses restricted diagnostics and a
small candidate budget.
Under \textsc{Full}, all configured route-compatible generation and
evaluation operations are available.
The implementation records the provisional route, executed route, downgrade
reason, exposed candidates, prescreening decisions, and incurred planning
cost.

\subsection*{B.2 Causal--Operational Valuation}

For each \(\omega\in\overline\Omega_t\), let
\(\widehat P_t(\omega)\),
\(\widehat C_t(\omega)\),
\(\widehat R_t(\omega)\),
\(\widehat I_t(\omega)\),
\(\widehat B_t(\omega)\), and
\(\widehat S_t(\omega\mid\chi_t)\)
denote estimated task progress, execution cost, operational risk,
information gain, robustness, and workflow safety, respectively.
All available signals are calibrated to a common bounded range before
aggregation.
A signal unavailable for a benchmark is omitted rather than assigned an
arbitrary value.

Define the shared progress--risk--safety score and the full-route
information--robustness bonus as
\begin{align}
\widehat Q_t(\omega)
&=
\alpha_{\mathrm{prog}}\widehat P_t(\omega)
-
\alpha_{\mathrm{risk}}\widehat R_t(\omega)
+
\alpha_{\mathrm{safe}}
\widehat S_t(\omega\mid\chi_t),
\label{eq:appendix_shared_operational_score}
\\
\widehat H_t(\omega)
&=
\alpha_{\mathrm{info}}\widehat I_t(\omega)
+
\alpha_{\mathrm{rob}}\widehat B_t(\omega).
\label{eq:appendix_full_operational_bonus}
\end{align}
The route-conditioned operational utility is
\begin{equation}
\label{eq:appendix_operational_utility}
\resizebox{\columnwidth}{!}{$
\begin{aligned}
\widehat U_{\mathrm{op}}
\left(
\omega\mid\chi_t,z_t
\right)
=
\begin{cases}
0,
&
z_t=\textsc{Skip},
\\
\widehat Q_t(\omega)
-
\alpha_{\mathrm{cost}}^{\mathrm{light}}
\widehat C_t(\omega),
&
z_t=\textsc{Light},
\\
\widehat Q_t(\omega)
-
\alpha_{\mathrm{cost}}^{\mathrm{full}}
\widehat C_t(\omega)
+
\widehat H_t(\omega),
&
z_t=\textsc{Full}.
\end{cases}
\end{aligned}
$}
\end{equation}
All coefficients are nonnegative.
The light route omits information-gain and robustness estimation when
restricted diagnostics are sufficient, whereas the full route uses the
complete configured valuation.

The causal--operational utility is
\begin{equation}
\label{eq:appendix_arco_utility}
\widehat U_{\mathrm{ARCO}}
\left(
\omega
\mid
\chi_t,
z_t,
\widehat\Gamma_{\phi,t}
\right)
=
\widehat U_{\mathrm{op}}
\left(
\omega
\mid
\chi_t,
z_t
\right)
+
\eta_\Gamma
\widehat\Gamma_{\phi,t}(\omega),
\end{equation}
where \(\eta_\Gamma\geq0\) controls the contribution of estimated workflow
effect evidence.
The strongest exposed candidate is
\begin{equation}
\label{eq:appendix_best_candidate}
\omega_t^+
=
\arg\max_{\omega\in\overline\Omega_t}
\widehat U_{\mathrm{ARCO}}
\left(
\omega
\mid
\chi_t,
z_t,
\widehat\Gamma_{\phi,t}
\right).
\end{equation}
Its advantage over the factual workflow is
\begin{equation}
\label{eq:appendix_candidate_advantage}
\begin{aligned}
\Delta_t^{\mathrm{cf}}
={}&
\widehat U_{\mathrm{op}}
\left(
\omega_t^+
\mid
\chi_t,
z_t
\right)
-
\widehat U_{\mathrm{op}}
\left(
\omega_t^0
\mid
\chi_t,
z_t
\right)
\\
&+
\eta_\Gamma
\left[
\widehat\Gamma_{\phi,t}(\omega_t^+)
-
\widehat\Gamma_{\phi,t}(\omega_t^0)
\right].
\end{aligned}
\end{equation}
The estimator enforces
\[
\widehat\Gamma_{\phi,t}(\omega_t^0)=0.
\]

\paragraph{Implementation correspondence.}
Task-progress signals may be derived from newly acquired evidence,
intermediate environment states, test results, or predicted goal advancement.
Cost signals use expected or observed model tokens, planner calls, tool calls,
and latency.
Risk signals capture invalid actions, destructive operations, execution
errors, or uncertain state changes.
Information-gain signals estimate whether the candidate obtains evidence
absent from the factual path.
Robustness signals favor candidates that depend less on unverified
assumptions or fragile execution sequences.
The final score ranks only candidates in the route-exposed admissible set \(\overline\Omega_t\).

\subsection*{B.3 Conservative Workflow Authorization}

The CRE decision for the selected candidate is
\begin{equation}
\label{eq:appendix_cre_prediction}
y_t^{\mathrm{chg}}
=
\arg\max_{y\in\{\textsc{Keep},\textsc{Change}\}}
p_{\theta_c}
\left(
y
\mid
\chi_t,
\omega_t^0,
\omega_t^+,
\widehat\iota_t
\right).
\end{equation}
\textsc{Keep} retains the factual workflow, whereas \textsc{Change} provides
learned support for replacement by \(\omega_t^+\).
CRE and causal--operational valuation have complementary roles:
CRE determines whether replacement is contextually supported, while
\(\Delta_t^{\mathrm{cf}}\) determines whether the estimated improvement is
large enough to justify revision.

The authorization rule is
\begin{equation}
\label{eq:appendix_revision_authorization}
\omega_t^\star
=
\begin{cases}
\omega_t^+,
&
y_t^{\mathrm{chg}}=\textsc{Change}
\ \land\
\Delta_t^{\mathrm{cf}}
\geq
\delta_{\widehat\iota_t},
\\[1mm]
\omega_t^0,
&
\text{otherwise},
\end{cases}
\end{equation}
where \(\delta_{\widehat\iota_t}\geq0\) is the authorization margin for the
predicted intervention family.
Because \(\omega_t^+\) is selected from \(\overline\Omega_t\), admissibility,
safety, schema, task-compatibility, and budget checks have already been
satisfied before this rule is applied.

For implementation diagnostics, define
\begin{equation}
\resizebox{\columnwidth}{!}{$
\begin{aligned}
a_t^{\mathrm{sup}}
=
\mathbb I
\left[
y_t^{\mathrm{chg}}=\textsc{Change}
\right]
\mathbb I
\left[
\Delta_t^{\mathrm{cf}}
\geq
\delta_{\widehat\iota_t}
\right],
a_t^{\mathrm{feas}}
=
\mathbb I
\left[
\omega_t^+
\in
\overline\Omega_t
\right],
\end{aligned}
$}
\end{equation}
and
\[
a_t^{\mathrm{chg}}
=
a_t^{\mathrm{sup}}
a_t^{\mathrm{feas}}.
\]
The learned CRE output and final authorization result are logged separately,
allowing prediction errors to be distinguished from constraint-triggered
fallback.
If either support or feasibility fails, ARCO retains \(\omega_t^0\).

\paragraph{Scope of workflow optimality.}
ARCO does not claim global optimization over the complete admissible workflow
space \(\mathcal A_t\).
Instead,
\[
\omega_t^+
=
\arg\max_{\omega\in\overline\Omega_t}
\widehat U_{\mathrm{ARCO}}
\left(
\omega
\mid
\chi_t,
z_t,
\widehat\Gamma_{\phi,t}
\right)
\]
is optimal only within the route-exposed admissible set.
CHILL-Harness executes this candidate when authorized and otherwise retains
the factual workflow.

\subsection*{B.4 Evidence-Grounded Completion and Answer Synthesis}

The CAE prediction is
\begin{equation}
\label{eq:appendix_cae_prediction}
\widehat b_t
=
\arg\max_{b\in\{0,1\}}
p_{\theta_e}
\left(
b
\mid
\chi_t,
o_t,
h_t
\right).
\end{equation}
The value \(\widehat b_t=1\) indicates learned completion support, whereas
\(\widehat b_t=0\) indicates that completion is unsupported by the CAE head.

An explicit task verifier produces
\begin{equation}
\label{eq:appendix_completion_evidence}
e_t^{\mathrm{comp}}
=
\mathbb I
\left[
\mathcal V_t
\left(
\chi_t,
o_t,
h_t
\right)
=1
\right],
\end{equation}
where \(\mathcal V_t\) is the task-specific completion verifier.
Termination is authorized by
\begin{equation}
\label{eq:appendix_completion_authorization}
d_t^{\mathrm{term}}
=
\widehat b_t
e_t^{\mathrm{comp}}.
\end{equation}
Thus, termination requires both learned completion support and explicit task
evidence.

The verifier may decompose completion evidence into strong, weak, and
contradictory signals:
\[
e_t^{\mathrm{comp}}
=
\mathbb I
\left[
E_t^{\mathrm{strong}}=1
\ \lor\
\left(
E_t^{\mathrm{weak}}=1
\ \land\
E_t^{\mathrm{fail}}=0
\right)
\right].
\]
Strong evidence includes passing tests, successful builds, accepted
solutions, or explicit environment success.
Weak evidence includes a valid final answer or task-completion signal,
whereas traceback, assertion, permission, and execution failures constitute
contradictory evidence.

\paragraph{Implementation correspondence.}
The CAE head does not directly invoke the terminal action; its output is
passed to the task-specific verifier.
Depending on the benchmark, the verifier checks a required answer format,
successful tests or evaluator results, the existence and validity of a
required artifact, an environment-provided success state, or another explicit
completion condition.
Predicted futility, budget exhaustion, repeated failure, or lack of progress
does not set \(e_t^{\mathrm{comp}}=1\).
Such signals may instead redirect execution toward \textsc{AnswerSynthesis}
or another workflow adaptation.
Therefore,
\[
\textsc{AnswerSynthesis}
\neq
\textsc{Complete}.
\]
The former is a nonterminal adaptation; the latter is an authorized terminal
decision.
The implementation records the CAE output, verifier result, supporting
evidence, synthesis decision, and final terminal action.


\setcounter{equation}{0}
\renewcommand{\theequation}{C\arabic{equation}}

\section*{Appendix C: Success-Preserving Learning and Complete Inference}

This appendix specifies the trajectory-level objective, the separated
supervision scheme, local authorization constraints, offline paired-effect
collection, and the complete CHILL-Harness inference procedure.

\subsection*{C.1 Success-Preserving Trajectory Objective}

The learnable CIEL parameter set is
\[
\Theta
=
\left\{
\phi,
\theta_u,
\theta_p,
\theta_c,
\theta_e
\right\}.
\]
ARCO introduces no separate prediction parameters in the current
formulation.
It realizes CIEL outputs through candidate generation, valuation, filtering,
and rule-based authorization with calibrated weights and thresholds.

Let \(\mathcal K\) denote the resource channels measured for a benchmark.
The trajectory-level execution cost is
\begin{equation}
\label{eq:appendix_execution_cost}
\mathcal J_{\mathrm{eff}}(\tau)
=
\sum_{k\in\mathcal K}
\lambda_k C_k(\tau),
\end{equation}
where \(C_k(\tau)\) is the amount of resource \(k\) consumed by trajectory
\(\tau\), and \(\lambda_k\geq0\) is its weight.
For the reported experiments, the principal resource channels are
\[
C_{\mathrm{tok}}(\tau)
=
C_{\mathrm{main}}^{\mathrm{in}}(\tau)
+
C_{\mathrm{main}}^{\mathrm{out}}(\tau)
+
C_{\mathrm{plan}}^{\mathrm{in}}(\tau)
+
C_{\mathrm{plan}}^{\mathrm{out}}(\tau),
\]
and
\begin{equation}
\resizebox{\columnwidth}{!}{$
\begin{aligned}
C_{\mathrm{time}}(\tau)
=
T_{\mathrm{main}}(\tau)
+
T_{\mathrm{plan}}(\tau)
+
T_{\mathrm{tool}}(\tau)
+
T_{\mathrm{env}}(\tau)
+
T_{\mathrm{verify}}(\tau).
\end{aligned}
$}
\end{equation}
The reported total token count includes main-agent and counterfactual-planner
calls made during benchmark inference.
Offline paired-replay collection is a training-data construction procedure
and is not included in test-time token or runtime measurements; its cost is
tracked separately.

The global objective is
\begin{equation}
\label{eq:appendix_global_objective}
\begin{aligned}
\min_{\Theta}\quad&
\mathbb E_{\tau\sim\pi_{\mathrm{CHILL},\Theta}}
\left[
\mathcal J_{\mathrm{eff}}(\tau)
\right]
\\
\text{s.t.}\quad&
\operatorname{PassRate}
\left(
\pi_{\mathrm{CHILL},\Theta}
\right)
\geq
\operatorname{PassRate}
\left(
\pi_{\mathrm{ref}}
\right)
-
\epsilon_{\mathrm{pass}}.
\end{aligned}
\end{equation}
Here, \(\pi_{\mathrm{CHILL},\Theta}\) is the execution policy induced by
CIEL and ARCO; \(\pi_{\mathrm{ref}}\) is the matched reference harness;
\(\operatorname{PassRate}(\pi)\) is the fraction of evaluated tasks completed
successfully by policy \(\pi\); and \(\epsilon_{\mathrm{pass}}\geq0\) is the
allowed aggregate pass-rate degradation.
Success is determined by the benchmark evaluator rather than by the harness
completion prediction.

\subsection*{C.2 Trajectory Supervision and Decision-Target Construction}

Let
\[
Y_\tau\in\{0,1\}
\quad\text{and}\quad
Y_{\mathrm{ref}}\in\{0,1\}
\]
indicate whether the CHILL-Harness and matched reference trajectories
succeed, respectively.
The trajectory supervision score is
\begin{equation}
\label{eq:appendix_trajectory_loss}
\resizebox{\columnwidth}{!}{$
\begin{aligned}
\mathcal L_{\mathrm{traj}}(\tau)
=
\mathcal J_{\mathrm{eff}}(\tau)
+
\lambda_{\mathrm{fail}}
\left(
1-Y_\tau
\right)
+
\lambda_{\mathrm{break}}
\mathbb I
\left[
Y_{\mathrm{ref}}=1
\ \land\
Y_\tau=0
\right].
\end{aligned}
$}
\end{equation}
The first term measures execution cost, the second penalizes task failure,
and the third penalizes adaptations that break a successful reference
trajectory.
The coefficients \(\lambda_{\mathrm{fail}},\lambda_{\mathrm{break}}\geq0\)
control the two failure penalties.

\paragraph{Separated supervision correspondence.}
The trajectory score is not differentiated through external tools or the
environment.
Instead, it determines valid samples, confidence weights, and weak decision
targets:
\[
\mathcal L_{\mathrm{traj}}
\longrightarrow
\left(
z_t^*,
y_t^{\mathrm{chg},*},
b_t^*
\right).
\]
Workflow-effect supervision is constructed separately:
\[
\mathcal D_{\mathrm{pair}}
\longrightarrow
\widetilde\Gamma_t^{\mathrm{pair}}(\omega)
\longrightarrow
\widehat\Gamma_\phi
\longrightarrow
\iota_t^*.
\]
Thus, retrospective trajectory evidence does not directly define or
supervise the identified workflow-effect target.
CDE targets select the least costly route preserving decision quality;
CRE targets determine whether the selected candidate should replace the
factual workflow; and CAE targets determine whether completion is already
supported by evidence.

\subsection*{C.3 Local Authorization Constraints}

Using the unified safety score from Appendix B, the admissible workflow set is
\begin{equation}
\label{eq:appendix_admissible_set}
\begin{aligned}
\mathcal A_t
=
\{\omega:\;&
\widehat S_t
\left(
\omega
\mid
\chi_t
\right)
\geq
\xi_t
\\
&\land\
\operatorname{SchemaValid}(\omega)
\\
&\land\
\operatorname{TaskCompatible}
\left(
\omega;
\chi_t
\right)
\\
&\land\
\operatorname{WithinBudget}
\left(
\omega;
\chi_t
\right)
\\
&\land\
\neg
\operatorname{Repeated}_t(\omega)
\}.
\end{aligned}
\end{equation}
Here, \(\widehat S_t(\omega\mid\chi_t)\) is the workflow-safety score and
\(\xi_t\) its minimum threshold;
\(\operatorname{SchemaValid}\) checks workflow and tool-call structure;
\(\operatorname{TaskCompatible}\) checks that proposed operations are
permitted in the current task and environment;
\(\operatorname{WithinBudget}\) checks remaining model, planning, tool,
revision, and execution budgets; and
\(\operatorname{Repeated}_t\) identifies duplicate or repeatedly ineffective
workflows.

Each generated candidate is parsed and passed through these checks before
valuation and authorization.
Rejected candidates are removed from
\[
\overline\Omega_t
=
\Omega_t\cap\mathcal A_t,
\]
and their rejection reasons are stored in the execution trace.
Revision satisfies
\[
\omega_t^\star=\omega_t^+
\Longrightarrow
\left[
y_t^{\mathrm{chg}}=\textsc{Change}
\ \land\
\Delta_t^{\mathrm{cf}}
\geq
\delta_{\widehat\iota_t}
\right],
\]
and completion satisfies
\[
d_t^{\mathrm{term}}=1
\Longrightarrow
\left[
\widehat b_t=1
\ \land\
e_t^{\mathrm{comp}}=1
\right].
\]
Together, these constraints ensure
\[
\omega_t^\star
\in
\overline\Omega_t
\subseteq
\mathcal A_t
\]
and prevent unsupported revision or premature termination.

\subsection*{C.4 Complete Inference, Offline Pairing, and Logging}

\begin{algorithm}[H]
\caption{Complete CHILL-Harness Inference}
\label{alg:appendix_chill}
\begin{algorithmic}[1]

\STATE \textbf{Input}:
\(\chi_t,\omega_t^0,o_t,h_t\);
\textbf{Output}:
\(\omega_t^\star,d_t^{\mathrm{term}}\)

\STATE Predict
\[
\widehat\iota_t
=
\arg\max_{\iota\in\mathcal I_t^{\mathrm{wf}}}
p_{\theta_u}
\left(
\iota
\mid
\chi_t,
\omega_t^0
\right)
\]

\STATE Predict the provisional route
\[
\widetilde z_t
=
\arg\max_{z\in\mathcal Z}
p_{\theta_p}
\left(
z
\mid
\chi_t,
\omega_t^0,
\widehat\iota_t
\right)
\]
and select the highest feasible route \(z_t\preceq\widetilde z_t\)

\STATE Generate the raw route-conditioned candidate set
\(\Omega_t^{\mathrm{raw}}\) using
Equation~\eqref{eq:appendix_route_candidate_set}

\STATE Prescreen candidates using
Equation~\eqref{eq:appendix_prescreened_candidates} and obtain \(\Omega_t\)

\STATE Filter admissible candidates:
\[
\overline\Omega_t
=
\Omega_t\cap\mathcal A_t
\]

\STATE For each \(\omega\in\overline\Omega_t\), estimate
\[
\widehat\Gamma_{\phi,t}(\omega)
=
\widehat\Gamma_\phi
\left(
\omega;
\chi_t,
\omega_t^0
\right)
\]
and compute
\(\widehat U_{\mathrm{ARCO}}
(\omega\mid\chi_t,z_t,\widehat\Gamma_{\phi,t})\)

\STATE Select \(\omega_t^+\) and compute \(\Delta_t^{\mathrm{cf}}\)
using Equations~\eqref{eq:appendix_best_candidate} and
\eqref{eq:appendix_candidate_advantage}

\STATE Predict CRE support:
\[
y_t^{\mathrm{chg}}
=
\arg\max_{y\in\{\textsc{Keep},\textsc{Change}\}}
p_{\theta_c}
\left(
y
\mid
\chi_t,
\omega_t^0,
\omega_t^+,
\widehat\iota_t
\right)
\]

\STATE Authorize \(\omega_t^\star\) using
Equation~\eqref{eq:appendix_revision_authorization}

\STATE Compute
\[
\widehat b_t
=
\arg\max_{b\in\{0,1\}}
p_{\theta_e}
\left(
b
\mid
\chi_t,
o_t,
h_t
\right)
\]
and
\[
d_t^{\mathrm{term}}
=
\widehat b_t
\mathbb I
\left[
\mathcal V_t
\left(
\chi_t,
o_t,
h_t
\right)
=1
\right]
\]

\IF{\(d_t^{\mathrm{term}}=1\)}
    \STATE Invoke the task-terminal action
\ELSE
    \STATE Execute \(\omega_t^\star\)
\ENDIF

\STATE Log decisions, candidate evaluations, rejection reasons, execution
costs, observations, errors, and task outcomes

\end{algorithmic}
\end{algorithm}

The inference record contains at least
\begin{equation}
\resizebox{\columnwidth}{!}{$
\begin{aligned}
\left(
\chi_t,
\omega_t^0,
\widehat\iota_t,
\widetilde z_t,
z_t,
\Omega_t^{\mathrm{raw}},
\Omega_t,
\overline\Omega_t,
\omega_t^+,
\Delta_t^{\mathrm{cf}},
y_t^{\mathrm{chg}},
\omega_t^\star,
\widehat b_t,
e_t^{\mathrm{comp}},
d_t^{\mathrm{term}}
\right),
\end{aligned}
$}
\end{equation}
together with tool calls, observations, rejection reasons, costs, errors, and
final task outcomes.

\paragraph{Offline paired-effect collection.}
Checkpointed paired replay is performed only during offline training-data
construction.
For each selected training checkpoint and counterfactual candidate, the
implementation restores the same checkpoint, executes the factual and
candidate branches under the common continuation and evaluation protocol,
computes Equation~\eqref{eq:appendix_paired_effect_target}, and adds the
resulting record to \(\mathcal D_{\mathrm{pair}}\).
CIEL parameters are then trained on the designated training split and frozen
before benchmark evaluation.
No paired replay, effect-model update, or decision-head update is performed on
evaluation tasks.

The complete learning and execution flow is
$
\text{offline paired execution}
\longrightarrow
\widetilde\Gamma_t^{\mathrm{pair}}
\longrightarrow
\widehat\Gamma_\phi
\longrightarrow
\iota_t^*,$
$\text{trajectory evidence}
\longrightarrow
\left(
z_t^*,
y_t^{\mathrm{chg},*},
b_t^*
\right),
$
$\text{frozen CIEL predictions}
\longrightarrow
\text{ARCO realization}
\longrightarrow
\text{authorized execution}.
$

\end{document}